%% file: sig-alternate.tex
\documentclass{sig-alternate}
\pdfoutput=1
\usepackage[export]{adjustbox}
\usepackage{forest}
\usepackage{algorithm}
\usepackage{algorithmic}
\usepackage{amsmath}
\usepackage{amsfonts}
\usepackage{amssymb}
\usepackage{graphicx}
\usepackage{tikz}
\usepackage{pgf}
\usepackage{multicol}
\usetikzlibrary{positioning}

\usepackage{amsthm}
\usepackage{times}
\usepackage{xspace}
\usepackage{multirow}
\usepackage[font={small,it}]{caption}
\usepackage[small, it, IT]{subfigure}

\newcommand{\stitle}[1]{\vspace{0.3em}\noindent\textbf{#1}}

\makeatletter
\def\@copyrightspace{\relax}
\makeatother

\newenvironment{denselist}{
    \begin{list}{\small{$\bullet$}}%
    {\setlength{\itemsep}{0ex} \setlength{\topsep}{0ex}
    \setlength{\parsep}{0pt} \setlength{\itemindent}{0pt}
    \setlength{\leftmargin}{1.5em}
    \setlength{\partopsep}{0pt}}}%
    {\end{list}}

\newcommand{\agp}[1]{\textcolor{magenta}{Aditya: #1}}

\newcommand{\orc}{{\sc Orchestra}\xspace}

\newcommand{\agpdelete}[1]{}
\newcommand{\extra}[1]{}

\newcommand{\papertext}[1]{#1}
\newcommand{\techreport}[1]{}
\newcommand{\pushedtoappendix}[1]{}

\newtheorem{definition}{Definition}[section]

\newtheorem{theorem}{Theorem}[section]

\newtheorem{problem}{Problem}[section]

\newtheorem{lemma}{Lemma}[section]

\begin{document}
%
\conferenceinfo{WOODSTOCK}{'97 El Paso, Texas USA}


\title{{\ttlit It's just a matter of perspective(s)}: \\ Crowd-Powered Consensus Organization of Corpora}

\numberofauthors{1}
\author{
\alignauthor Ayush Jain, Joon Young Seo, $^\dagger$Karan Goel, Andrew Kuznetsov \\ Aditya Parameswaran, Hari Sundaram\\
\affaddr{University of Illinois, Urbana Champaign, $^\dagger$IIT Delhi} \\
\affaddr{\{ajain42, jmseo2, akuznet2, adityagp, hs1\}@illinois.edu; $^\dagger$kgoel93@gmail.com}
}

\maketitle
\begin{abstract}
We study the problem of 
organizing a collection of objects---images, videos---into clusters,
using crowdsourcing. 
Th\-is problem is notoriously hard for computers to do automatically,
and even with crowd workers, is challenging to orchestrate:
{\em (a)} workers may cluster based on different latent hierarchies or perspectives;
{\em (b)} workers may cluster
at different granularities even when clustering using the same perspective; 
and {\em (c)} workers may only see a small portion of the objects when deciding
how to cluster them (and therefore have limited understanding
of the ``big picture'').
We develop cost-efficient, accurate algorithms for identifying 
the consensus organization (i.e., the organizing perspective most
workers prefer to employ), and incorporate these algorithms
into a cost-effective workflow for organizing a collection of objects,
termed \orc.
We compare our algorithms with other algorithms for clustering,
on a variety of real-world datasets, and demonstrate that
\orc organizes items better and at significantly lower costs.
\end{abstract}




\section{Introduction}\label{sec:intro}
\input{intro}
\input{prelim_worker.tex}

\input{creating_hierarchies.tex}\label{sec:algorithm}
\section{Experiments}

\input{experiments.tex}

\input{related_work.tex}
\section{Conclusions and Future Work}
We described \orc, our approach to perform 
consensus organization of corpora using the crowd.
We developed techniques for identifying 
maximum likelihood frontiers, for issuing
additional questions from the crowd, ensuring
that the eventual frontiers have high coverage,
and combining information across different crowd answers.
We demonstrated the benefits of \orc versus other
crowd-clustering schemes on three datasets
with different characteristics.
The organizations returned by \orc are higher
quality (up to 24\% improvement on VI) and are more cost effective 
(up to a reduction of $6\times$) than other
schemes.

We believe our paper raises a number of interesting
unanswered questions:
{\em (a)} Would it help to ask workers to describe, in words,
the clustering that they are using, and combine that information
with the hierarchy construction or merging algorithm?
Once we identify the maximum likelihood hierarchy, could
we ask workers to cluster on that hierarchy (in words)?
{\em (b)} Would it help at all to drill-down on certain nodes
in a given hierarchy by asking workers to only organize objects
that are known to be part of the concept corresponding to that node?
One drawback of this is that we may end up mixing hierarchies:
if we apply drill-down to a node containing triangles, we may
end up introducing size or color as the organizational principle 
at that point.
{\em (c)} We observed that often the hierarchies that we obtain (corresponding
to the cliques) may in fact share many clusterings: in such cases, we 
still just end up picking the largest clique. Would it be possible
to merge these hierarchies together, despite not being part of a single clique, by
using a more tolerant merging criteria---would that lead to any
benefits?
{\em (d)} Can we combine our algorithm with an automated scheme
that provides prior assessments of similarity using automatically
extracted features?
We plan to address these, and other questions in follow-up work.
{
\bibliographystyle{abbrv}
\bibliography{biblio}
}
\end{document}

%% file: intro.tex
\vspace{-5pt}
\begin{quote}
{\em ``Everything we hear is an opinion, not a fact. \\ Everything we see is a perspective, not the truth.''} \\ --- Marcus Aurelius, ca.~150 AD.
\end{quote}
\vspace{-5pt}

With the costs of storage rapidly decreasing, 
we have been amassing large volumes of images and
videos within our personal computers
and within shared file systems in organizations.
To be able to make effective use of these images and videos, 
it is essential to {\em organize}
them into clusters.
Unfortunately, automated schemes perform poorly at  
organization since they are not able to interpret
or understand content adequately.
Human beings, on the other hand, can easily organize
such content, but it is often impossible for
any single human worker to organize a large corpus. 
So we turn to crowdsourcing for organizing content.

Unfortunately, employing crowdsourcing
is rife with several issues, stemming from the fact that
there are often many correct ways of organizing complex content such as images.
To illustrate these issues (listed below), we asked 20 workers
on Amazon's Mechanical Turk to cluster
a stylized set of 25 images, 
where each image is a random combination
of (\textsc{shape}, \textsc{color}, \textsc{size}).
Workers were allowed to create as many clusters
as they wanted, and populate these
clusters with the 25 images.
We note that this is a simple experiment---we expect
real world corpora to be more complex.

\begin{denselist}
\item {\em Issue 1: Perspectives.} Human workers often organize items using 
distinct organizational perspectives,
rendering the answers or clusters obtained from different
workers incomparable, making it hard to 
combine opinions across workers.
For example,  in our experiment,
85\% of the workers chose to organize by 
\textsc{shape}, 10\% by \textsc{color}, 
and 5\% by \textsc{size}.

\item {\em Issue 2: Granularities.} Even within a single organizational perspective, 
workers often organize at 
different ``granularities''.
For instance, for workers that chose 
to organize based on \textsc{shape}, 
some chose to create the following clusters: \{{\tt Polygons}, {\tt Ellip\-ses}\}, 
while others chose to split the {\tt Polygons} cluster, giving 
us
\{{\tt Rectang\-les}, {\tt Trian\-gles}, {\tt Ellipses}\}. 
Consequently, the number of clusters given by the workers 
also varied drastically.

\item {\em Issue 3: Limited Understanding of the ``Big Picture''.} 
To limit cognitive load, workers 
can only cluster or organize a
small number of items at once, making it hard for them
to understand how the small set of items fits in with  
the rest.
For instance, if there were no triangles in the
set of 25 items given to a worker, they would
organize the items assuming that
triangles did not exist in the dataset, 
while that might not actually be true.

\end{denselist}
{\em  We 
focus on the problem of developing a cost-efficient robust workflow to
perform consensus organization
of large corpora}, one
that majority of the workers agree with.
In our experiment above, we found that majority
of the workers clustered on  {\sc shape},
and that would represent our consensus organizational
perspective.
Work from behavioral psychology
on {\em free classification}
has similarly demonstrated that humans
have a tendency to pick a specific (likely) organizational perspective,
while at the same time humans do adopt different 
perspectives~\cite{imai1965discriminability,regehr1995category,handel1972free,medin1987family,milton2004influence}.

Prior work has considered the problem of crowd clustering~\cite{gomes2011crowdclustering,yi2012semi,yi2012crowdclustering},
falling short in three ways:
{\em (a)} These papers do not take into account
the fact that different workers may organize using
different perspectives and at different granularities, 
leading to an organization that is sub-optimal
with mixed organizational perspectives.
{\em (b)} Prior work emphasizes the use of random sampling; 
however in the absence of any relationship between the subsets of samples 
that the workers see, randomized sampling is costly. 
Indeed, \cite{gomes2011crowdclustering} report in their paper that
they require each item to appear in 6 
random samples to ensure goodness of clustering,
making it impractical in terms of cost.
{\em (c)} These papers transform the clusters provided
by workers into votes on the similarity or dissimilarity of 
pairs of items, losing out on the overall clustering structure. 
This is because
the eventual goal of these papers is to 
recover pairwise similarity or dissimilarity information,
as opposed to finding a consensus organization. 
Due to these limitations, prior work can only organize
items appropriately if there is a single perspective
with no variable granularities
(which is not true even in our stylized example above
and certainly not true in real datasets).
Indeed, we find that on real datasets, 
their results are much worse.
We describe related work in more detail in Section~\ref{sec:related}.


Our workflow, termed \orc, instead uses workers to
repeatedly organize carefully selected groups of items.
Instead of decomposing the responses from workers
into pairwise comparisons, we
operate on them directly.
We develop algorithms to infer not just 
which organizational perspective a worker is clustering
using but also the granularity within.
We use these algorithms in conjunction with techniques
to identify the maximum likelihood granularity
in the maximum likelihood perspective,
assembled into a workflow 
for organization.

There are several challenges in assembling \orc. 
First, 
ensuring adequate coverage is hard---all clusters need to be well 
represented, even when individual workers may not see representatives from all clusters.
Second, 
it is not easy to identify if workers are clustering
on the same organizational perspective,
especially if they are using different granularities,
or combining granularities.
For instance, a worker may provide
triangles, squares, non-polygons as three clusters,
while another worker may provide
polygons, ellipses, circles as three clusters;
both these workers are using different granularities
on the same perspective.
Third, 
once we identify that workers are indeed clustering using
the same perspective, 
it is not trivial to combine information across workers.
In our example given previously, no two clusters provided
by workers are alike, making it challenging to combine information across them.
Fourth, 
combining or relating information across workers is exacerbated by the
fact that different workers may be clustering different sets of items;
we need to identify common ``pivots'' that can help us relate
clusters across workers on different sets of items. 
Last,
assembling repeated worker clusterings into a cost-effective
workflow, while setting the parameters
that control the workflow in a principled manner,
is yet another challenge.



Here is a list of technical contributions in this paper:
\begin{denselist}
\item We model the problem formally using {\em graph hierarchies} to capture
the notion of perspectives, and {\em frontiers} on the hierarchies to capture the 
notion of granularities. (Section~\ref{sec:prelim})
\item We design, \orc, a {\em robust, low-cost workflow for organization}
comprising the following algorithmic components:
\begin{denselist}
\item We develop techniques to map worker clusterings to hierarchies
(to identify worker perspectives), 
and formalize the identification of the consensus or the maximum likelihood hierarchy
as a {\sc Max-Clique} problem. (Section~\ref{sec:hierarchyConstruc})
\item We develop probabilistic techniques to ensure that our maximum likelihood hierarchy
has {\em adequate coverage} of the space of all concepts in the dataset. (Section~\ref{sec:Sampling Guarantee})
\item We develop the notion of a {\em kernel} to relate worker clusterings
on different samples of items to the maximum likelihood hierarchy. (Section~\ref{sec:generateSample})
\item We design techniques to {\em extend} the current maximum likelihood hierarchy by merging
worker responses on new items to the existing hierarchy. (Section~\ref{sec:mergingHierarchies})
\item We develop algorithms that operate {\em bottom-up} 
to identify the maximum likelihood frontier on the maximum likelihood 
hierarchy, which can then be leveraged for {\em categorization}, 
providing further savings on cost and improved accuracies. (Section~\ref{sec:categorize})
\end{denselist}
\item We further couple these algorithmic contributions with experiments on three real datasets
on Amazon's Mechanical Turk (Section~\ref{sec:exp}), 
and demonstrate that our techniques lead
to better quality clusterings, when compared both to prior work in this space, as well
more primitive versions of \orc.
\end{denselist}

%% file: prelim_worker.tex
\section{Preliminaries} \label{sec:prelim}
In this section we discuss some essential concepts and ideas. 
In Section~\ref{sec:datamodel}, 
we present a sequence of definitions that helps formalize 
the problem we address in this paper. 
In Section~\ref{sec:workerBehavior}, we describe our
model for worker behavior and our interfaces, 
and in Section~\ref{sec:flow}, we describe 
the \orc workflow at a high level.
Finally, in Section~\ref{sec:high-level}, we provide a 
breakdown of the clustering phase of \orc that
will be our focus in the next section.

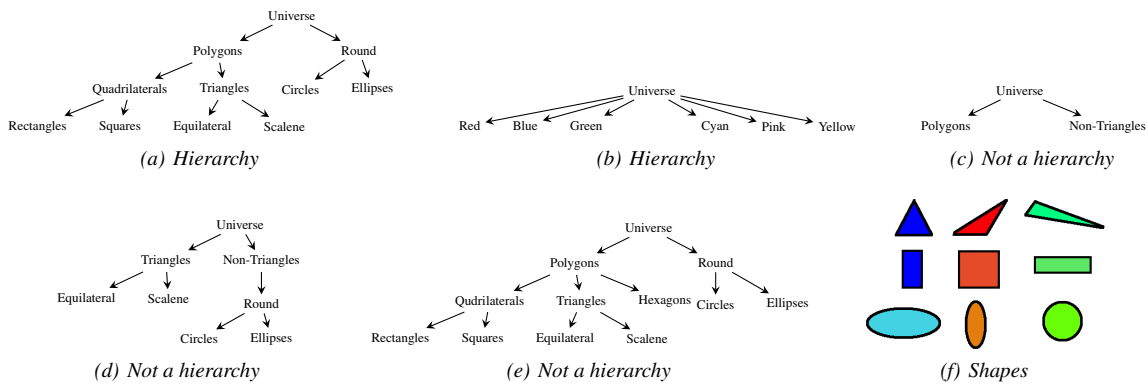
\begin{figure*}
\centering
\subfigure[][\label{fig:Hier1}Hierarchy]{
        
	\tiny{\begin{tikzpicture}[%
		->,
		>=stealth,
		node distance=0.3cm,
		pil/.style={
			->,
			thick,
			shorten =2pt,}
		]
		\node (univ) {Universe};
		\node[below left=of univ] (poly) {Polygons};
		\node[below left=of poly] (quad) {Quadrilaterals};
		\node[below right=of univ] (nonPoly){Round};
		\node[right=of quad] (tri) {Triangles};
		\node[below left=of quad] (rect) {Rectangles};
		\node[right=of rect] (sq) {Squares};
		\node[right=of sq] (eq) {Equilateral};
		\node[right=of eq] (scalene) {Scalene};
		\node[right=of tri] (circle) {Circles};
		\node[right=of circle] (ell) {Ellipses};
		\draw [->] (univ) to (poly);
		\draw [->] (univ) to (nonPoly);
		\draw [->] (poly) to (quad);
		\draw [->] (poly) to (tri);
		\draw [->] (nonPoly) to (circle);
		\draw [->] (nonPoly) to (ell);
		\draw [->] (quad) to (rect);
		\draw [->] (quad) to (sq);
		\draw [->] (tri) to (eq);
		\draw [->] (tri) to (scalene);
	\end{tikzpicture}}
	
	   }%
	   \qquad
	   \subfigure[][\label{fig:Hier2}Hierarchy]{
        
	\tiny{\begin{tikzpicture}[%
		->,
		>=stealth,
		node distance=0.3cm,
		pil/.style={
			->,
			thick,
			shorten =2pt,}
		]
		\node (univ) {Universe};
		\node[below right=of univ] (cyan) {Cyan};
		\node[below left=of univ] (green){Green};
		\node[left=of green] (blue) {Blue};		
		\node[left=of blue] (red) {Red};
		\node[right=of cyan] (pink) {Pink};
		\node[right=of pink] (yellow) {Yellow};
		\draw [->] (univ) to (red);
		\draw [->] (univ) to (blue);
		\draw [->] (univ) to (green);
		\draw [->] (univ) to (cyan);
		\draw [->] (univ) to (pink);
		\draw [->] (univ) to (yellow);
	\end{tikzpicture}}
	
	   }%
	   		   \qquad
	   \subfigure[][\label{fig:notHier2}Not a hierarchy]{
	\tiny{\begin{tikzpicture}[%
		->,
		>=stealth,
		node distance=0.3cm,
		pil/.style={
			->,
			thick,
			shorten =2pt,}
		]
		\node (univ) {Universe};
		\node[below left=of univ] (poly) {Polygons};
		\node[below right=of univ] (nonTri) {Non-Triangles};
		\draw [->] (univ) to (poly);
		\draw [->] (univ) to (nonTri);
	\end{tikzpicture}}
	}
	   \qquad

	   \subfigure[][\label{fig:notHier1}Not a hierarchy]{
	\tiny{\begin{tikzpicture}[%
		->,
		>=stealth,
		node distance=0.3cm,
		pil/.style={
			->,
			thick,
			shorten =2pt,}
		]
		\node (univ) {Universe};
		\node[below left=of univ] (tri) {Triangles};
		\node[right=of tri] (nonTri) {Non-Triangles};
		\node[below=of nonTri] (nonPoly){Round};
		\node[below left=of tri] (eq) {Equilateral};
		\node[right=of eq] (scalene) {Scalene};
		\node[below left=of nonPoly] (circle) {Circles};
		\node[right=of circle] (ell) {Ellipses};
		\draw [->] (univ) to (poly);
		\draw [->] (univ) to (nonTri);
		\draw [->] (nonTri) to (nonPoly);
		\draw [->] (nonPoly) to (circle);
		\draw [->] (nonPoly) to (ell);
		\draw [->] (tri) to (eq);
		\draw [->] (tri) to (scalene);
	\end{tikzpicture}}
	}
	\qquad
	\subfigure[][\label{fig:notHier3}Not a hierarchy]{
	\tiny{\begin{tikzpicture}[%
		->,
		>=stealth,
		node distance=0.3cm,
		pil/.style={
			->,
			thick,
			shorten =2pt,}
		]
		\node (univ) {Universe};
		\node[below left=of univ] (poly) {Polygons};
		\node[below left=of poly] (quad) {Qudrilaterals};
		\node[below right=of univ] (nonPoly){Round};
		\node[right=of quad] (tri) {Triangles};
		\node[right=of tri] (hexagons) {Hexagons};
		\node[below left=of quad] (rect) {Rectangles};
		\node[right=of rect] (sq) {Squares};
		\node[right=of sq] (eq) {Equilateral};
		\node[right=of eq] (scalene) {Scalene};
		\node[below=of nonPoly] (circle) {Circles};
		\node[right=of circle] (ell) {Ellipses};
		\draw [->] (univ) to (poly);
		\draw [->] (univ) to (nonPoly);
		\draw [->] (poly) to (quad);
		\draw [->] (poly) to (tri);
		\draw [->] (poly) to (hexagons);
		\draw [->] (nonPoly) to (circle);
		\draw [->] (nonPoly) to (ell);
		\draw [->] (quad) to (rect);
		\draw [->] (quad) to (sq);
		\draw [->] (tri) to (eq);
		\draw [->] (tri) to (scalene);
	\end{tikzpicture}}
	   }%
	   	   \qquad
	\subfigure[][Shapes]{\label{fig:shapes}\includegraphics[scale=0.2]{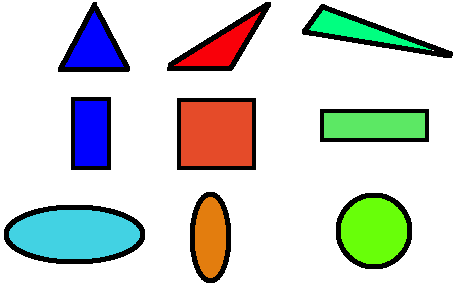}}
	\vspace{-10pt}
	\caption{\label{fig:hierarchies}(a) -- (e): Concept trees for the clustering example shown in Figure~\ref{fig:clusterinterface} --- (a) and (b) are hierarchies; (c) is not a hierarchy since it violates (3) in Definition~\ref{def:hierarchy} --- quadrilaterals in the dataset are instances of both children of {\tt Universe}; (d) is not a hierarchy since it violates (2) --- quadrilaterals in the dataset are instances of {\tt Non-Triangles} but not of any children;  (e) is not a hierarchy --- {\tt Hexagons} is a superfluous concept for this dataset. (f) Some examples of items in our Shapes dataset, which we use as a running example in this paper}
	\vspace{-10pt}
\end{figure*}

\subsection{Data Model}\label{sec:datamodel}
In this subsection, we provide a series of definitions related to four ideas: clusterings, hierarchies, frontiers and complete frontiers. First, we begin with a formal definition of clustering. 


\begin{definition}[{\bf Clustering}]
Given a set of items $\mathcal{D}$, a clustering is a partitioning of $\mathcal{D}$ into clusters $C_{1},\dots, C_{k}$ such that,
\vspace{-5pt}
\begin{multicols}{2}
\begin{enumerate}
\item[(1)] $C_{i} \cap C_{j} = \emptyset \\ \forall \, i \neq j \, \in \, \{1,\dots,k\}$
\item[(2)] $\bigcup_{i = 1} ^{k} C_{i} = \mathcal{D}$
\end{enumerate}
\end{multicols}

\end{definition}
\vspace{-5pt}
Every cluster in a clustering (and by consequence any set of items) can be associated with an {\it underlying latent concept}. Intuitively, a concept is a description that is satisfied by each item in a cluster. For example, in Figure~\ref{fig:shapes}, the clusters---from top to bottom, one corresponding to each row---represent the concepts {\tt Triangles}, {\tt Quadrilaterals} and {\tt Ellipses}. Formally, a concept describes the set of common attributes shared by all items in a cluster. We say that the items in a cluster are {\it instances} of its latent concept. Anything that holds true for a concept, also holds true for the cluster that it represents. 


Concepts may have subset-superset relationships among them
. Formally, we say that concept $B$ {\it generalizes} concept $A$ (denoted $B \succ A$) if every item in $\mathcal{D}$ that is an instance of $A$ is also an instance of $B$. For example, the concept {\tt Quadrilaterals} generalizes {\tt Rectangles}. We introduce the concept {\tt Universe}, which describes any item in the corpus $\mathcal{D}$. By definition, {\tt Universe} generalizes every concept associated with any subset of $\mathcal{D}$.

We can organize concepts based on the {\it generalize} relationship into a rooted concept tree.  We call this concept tree a {\it hierarchy}.

\begin{definition}[{\bf Hierarchy}]
\label{def:hierarchy}
For the set of items $\mathcal{D}$, a {hierarchy} $\mathcal{T}(\mathcal{D})$ is a rooted concept tree where
\vspace{-2pt}
\begin{enumerate}\itemsep -2pt
\item[(1)] A concept $A \in \mathcal{T}(\mathcal{D})$ is a parent of another concept $B \in \mathcal{T}(\mathcal{D})$ if $A \succ B$ and there exists no $C \in \mathcal{T}(\mathcal{D})$ such that $A \succ C$ and $C \succ B$
\item[(2)] {\tt Universe} is the root node of $\mathcal{T}$
\item[(3)] Every instance of $C \in \mathcal{T}(\mathcal{D})$ is also an instance of exactly one of its children in $\mathcal{T}$ 
\item[(4)] For every $C \in \mathcal{T}(\mathcal{D})$, at least one item in $\mathcal{D}$ is an instance of $C$.
\end{enumerate}
\end{definition}


\noindent Intuitively, a hierarchy is a concept tree in which every item of $\mathcal{D}$ can be assigned to exactly one of the leaf nodes (and consequently all of its ancestors), and no leaf node is empty. Multiple datasets may have the same hierarchy, and a dataset may be representable by multiple hierarchies.  

Note that while a hierarchy is defined in terms of concepts, each concept can be replaced by the cluster that it describes, to get a hierarchy of clusters, built on the subset relation. We will treat these hierarchies as equivalent.

Figure~\ref{fig:hierarchies} shows some concept trees for the Shapes dataset items shown in Figure~\ref{fig:clusterinterface}. Figures~\ref{fig:Hier1} and~\ref{fig:Hier2} are hierarchies as every item in the dataset can be assigned to one of the leaf nodes. Other trees, shown in Figure~\ref{fig:notHier2}, \ref{fig:notHier1} and \ref{fig:notHier3}, are not hierarchies. Figure~\ref{fig:notHier2} is not a hierarchy because the concepts {\tt Polygons} and {\tt Non-Triangles} are not disjoint. Rectangles in the dataset are instances of both concepts and cannot lie in exactly one of them. In~\ref{fig:notHier1}, the concept {\tt Round} does not cover all instances of its parent concept {\tt Non-Triangles}. The dataset has a {\tt Quadrilaterals} concept in addition to {\tt Round}. Figure~\ref{fig:notHier3} is also not a hierarchy as there are no instances of {\tt Hexagons} in the dataset.

We now describe a method to find the hierarchy corresponding to any subset of $\mathcal{D}$, when $\mathcal{T}(\mathcal{D})$ is given. Let there be some set of items $\mathcal{S} \subseteq \mathcal{D}$ associated with $C \in \mathcal{T}(\mathcal{D})$ such that every item in $\mathcal{S}$ is an instance of $C$. $\mathcal{S}$ may or may not contain every instance of $C$. Consider the subtree of $\mathcal{T}(\mathcal{D})$ rooted at $C$. If we enforce condition (2) and (4) in our definition of a hierarchy --- replacing $C$ by the {\tt Universe} placeholder, and dropping superfluous concept nodes in this subtree --- the resulting tree will be a hierarchy $\mathcal{T}(\mathcal{S})$. For instance, in Figure~\ref{fig:Hier1}, the subtree rooted at {\tt Polygons} is a hierarchy if $S$ is the set of all polygons in the dataset. If $S$ only contains squares and all triangles, then we would remove {\tt Rectangles} as it is now a superfluous concept, and the leftover tree would be a hierarchy. 
We now define the concept of a frontier.


\begin{definition}[{\bf Frontier}]
A frontier $F$ is a set of disjoint concepts $\{C_{1},\dots,C_{k}\}$ in a hierarchy $\mathcal{T}(\mathcal{D})$ such that: \\
$
\hphantom{fun stuff} \nexists \ i, j \ \in \ \{1,\dots , k\} :  C_{i} \succ C_{j}
$
\end{definition}
\noindent In words, a frontier is a set of disjoint concepts such that no two concepts in a frontier are connected by the \textit{generalizes} relationship. For the hierarchy shown in Figure~\ref{fig:Hier2}, \{{\tt Red}, {\tt Green}, {\tt Blue}\} forms a valid frontier. Since concepts in $F$ are disjoint, an item in $\mathcal{D}$ can be an instance of {\it atmost} one concept in $F$. 

\begin{definition}[{\bf Complete Frontier}]
A frontier $F$ in $\mathcal{T}(\mathcal{D})$ is said to be complete if \ \ 
$\bigcup_{i= 1}^k C_{i} = \text{{\tt Universe}}$
\end{definition}
\noindent In other words, $F$, it is said to be a complete frontier if every item in $\mathcal{D}$ is an instance of {\it exactly} one concept in $F$. For the hierarchy of Figure~\ref{fig:Hier2}, the frontier \{\texttt{Red}, \texttt{Blue}, \texttt{Green}\}, when expanded to \{\texttt{Red}, \texttt{Blue}, \texttt{Green}, \texttt{Cyan}, \texttt{Pink}, \texttt{Yellow}\} becomes complete as every item in the dataset is an instance of exactly one of these concepts. Similarly, for Figure~\ref{fig:Hier1}, \{{\tt Polygons},  {\tt Circles},  {\tt Ellipses}\}, \{{\tt Quadrilaterals},  {\tt Triangles},  {\tt Round}\}, \{{\tt Rectangles},  {\tt Squares},  {\tt Equilateral},  {\tt Scalene},  {\tt Circles},  \\{\tt Ellipses}\} are all complete frontiers. 

Notice the similarities in the definition of clustering and that of a complete frontier.  Just as a cluster operationalizes a concept, a clustering can be viewed as an operationalization of a complete frontier on a set of items. Thus, a complete frontier is associated with a clustering of the dataset.

\subsection{Interacting with Workers}
\label{sec:workerBehavior}

\begin{figure}[htbp]
\begin{center}
\includegraphics[width=0.7\columnwidth]{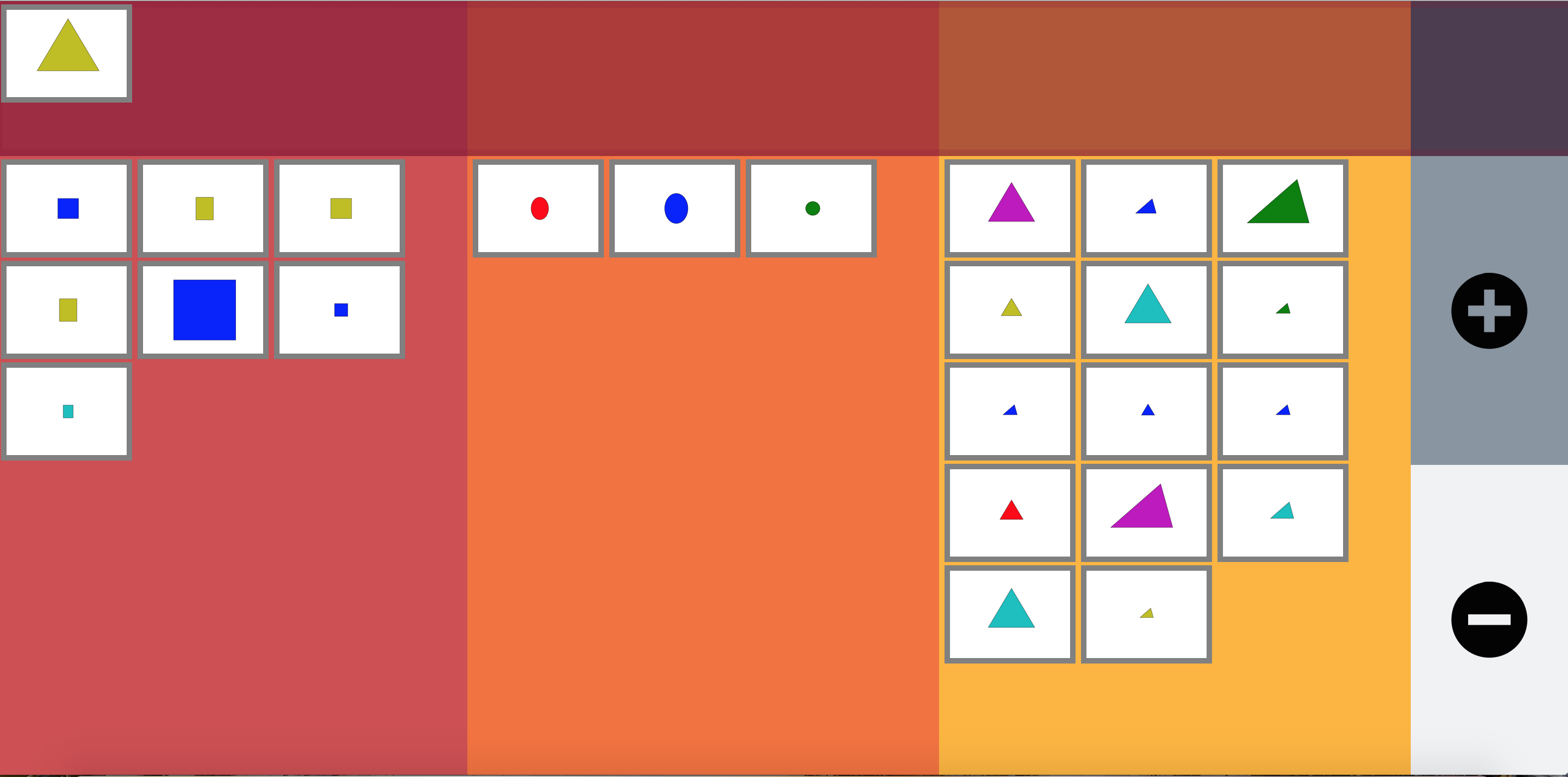}
\caption{Our clustering interface. In this example, workers are asked to organize shapes into multiple clusters. They can determine the number of clusters by using the `+' and the `-' buttons seen on the right.}
\label{fig:clusterinterface}
\end{center}
\vspace{-20pt}
\end{figure}

We use two interfaces to interact with workers. 
The first interface is a {\em clustering interface}.
Here, workers are presented with a carousel of items, which they can drag into as many clusters as they like. This interface allows us to generate partial clusterings for a small set of items.  See Figure~\ref{fig:clusterinterface} for an example worker session.

We model the response to this interface, resulting in a clustering, as a frontier in some latent, underlying hierarchy.
Different workers may have completely different latent hierarchies in mind; for instance, Figures~\ref{fig:Hier1} and~\ref{fig:Hier2} are both valid hierarchies for the data shown in Figure~\ref{fig:clusterinterface}.  Thus, the worker clustering process can be modeled as follows. First, given a subset $\mathcal{S} \in \mathcal{D}$, a worker picks some latent hierarchy $\mathcal{T}(\mathcal{S})$.  Then, the worker chooses  a complete frontier $F$ in $\mathcal{T}(\mathcal{S})$.  Notice that while $F$ is complete in $\mathcal{T}(\mathcal{S})$, it will not in general be complete in $\mathcal{T}(\mathcal{D})$. Finally, the output of the worker is the clustering of $\mathcal{S}$ associated with $F$.

We also use a {\em categorization interface}, which is similar to the clustering
interface except that a fixed number of clusters are shown, and each cluster is pre-populated
with a fixed set of items. Workers are asked to drag the new items into one of these existing clusters,
thereby categorizing them. 
In this case, workers no longer have the freedom to select their own latent
hierarchy for organization and must instead use the clustering already provided.

\subsection{Overall Workflow for {\large \orc}}\label{sec:flow}
\begin{figure}[t!]
\centering
\vspace{-10pt}
\hspace{-10pt}
\subfigure[Our Workflow]{\includegraphics[scale=0.3,valign=m]{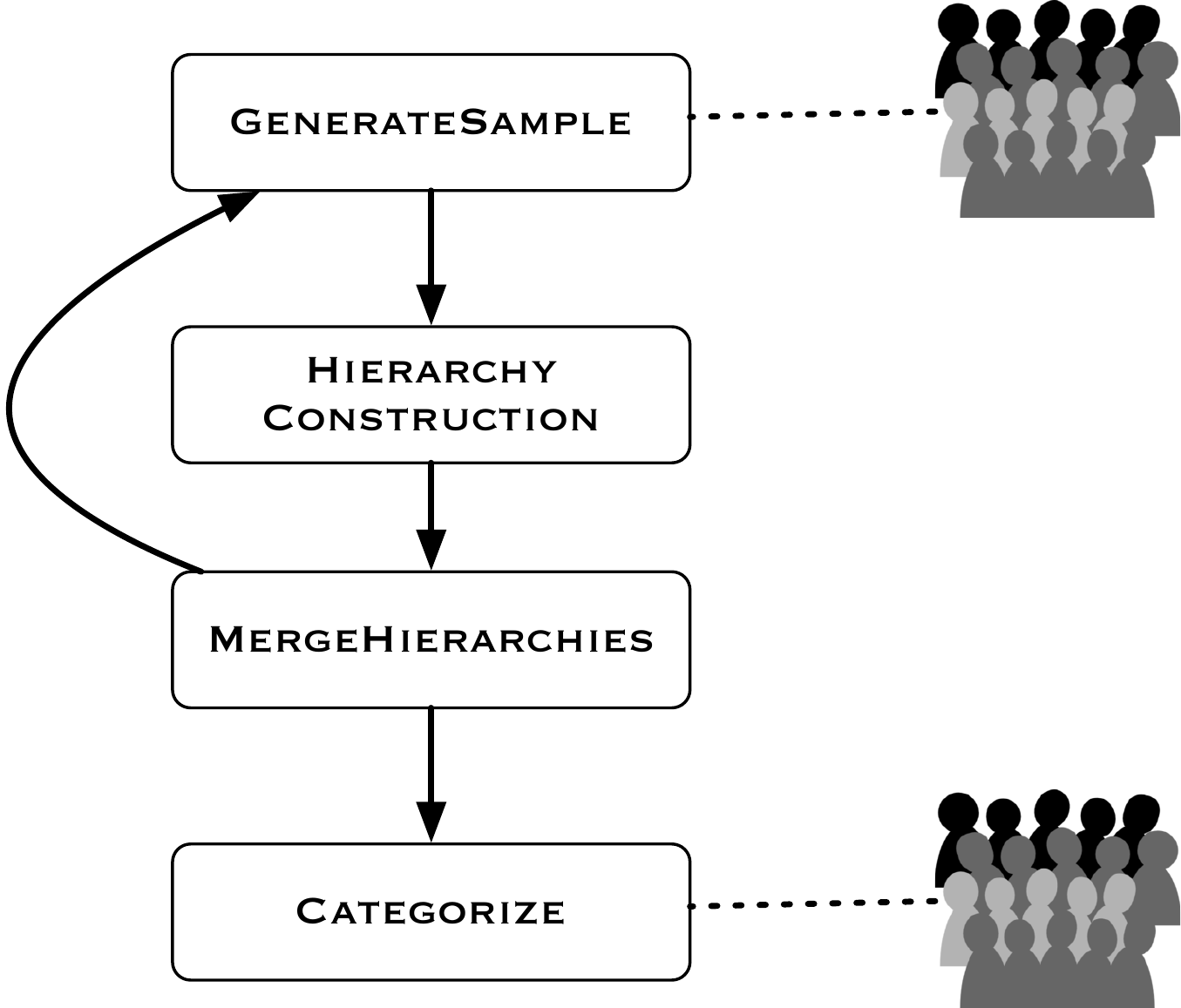} 
\label{fig:our-work}}%
\subfigure[Prior Work Workflow]{\vspace{-40pt}\includegraphics[scale=0.3,valign=m]{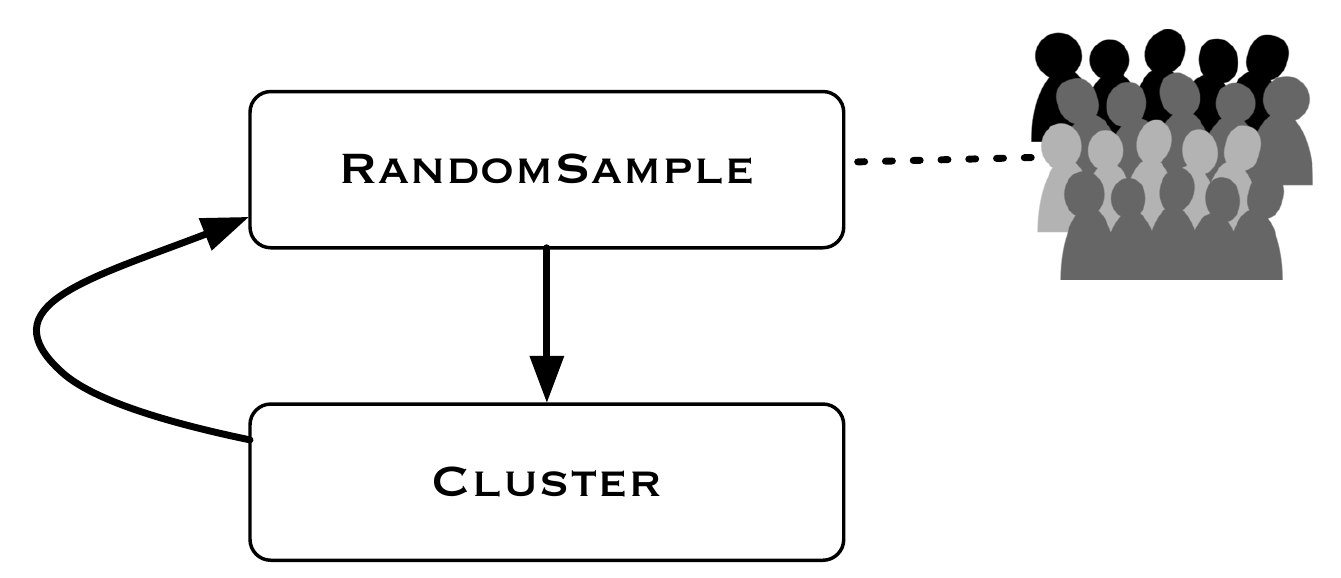}
\label{fig:prior-work}}
\vspace{-10pt}
\caption{Comparison of Workflows}
\vspace{-20pt}

\end{figure}

Our overall workflow comprises of two phases: the clustering phase and the categorization phase.
The clustering phase discovers a consensus organization of the data using just a small fraction of items from the corpus. 
Once the consensus set of clusters are determined, most of the items are then organized in the categorization phase, where we place items into clusters with which they share greatest similarity. Unlike previous work~\cite{gomes2011crowdclustering,yi2012crowdclustering}, we \emph{don't} make workers cluster every item in the dataset, which allows us to cut costs significantly. 
Also unlike previous work, \emph{we do not randomly sample} items in each iteration. Instead, we systematically pick some items that are already part of the hierarchy, so that new clusterings can be easily integrated into it.
 See Figures~\ref{fig:our-work} and~\ref{fig:prior-work}  for a graphical comparison between \textsc{Orchestra} and prior work. 
In Figure~\ref{fig:our-work}, the first three boxes refer to the clustering phase, 
while the last one refers to the categorization phase. 

The categorization phase is straightforward, with the only goal being to categorize
the remaining items in the dataset; categorization will be applied to the majority of the items. 
The transition from the clustering to the categorization phase will depend on the dataset complexity.
Our primary focus will be the clustering phase; we describe how it is broken down, next.

\subsection{Clustering Phase for {\large \orc}}\label{sec:high-level}

Given a dataset $\mathcal{D}$, the goal of the clustering phase is to recover the maximum likelihood latent hierarchy $\mathcal{T}_{ML}(\mathcal{D})$. This hierarchy has maximum likelihood in the sense that a worker clustering the entire dataset  would pick $\mathcal{T}_{ML}(\mathcal{D})$ as the latent organizational hierarchy with the highest probability.

We need to generate $\mathcal{T}_{ML}(\mathcal{D})$ across multiple samples of the dataset. This is because in any realistic setting with large datasets, workers  will cluster a dataset $\mathcal{S}$ where $\mathcal{S} \subset \mathcal{D}$, and indeed, in general it is likely that $|\mathcal{S}| \ll |\mathcal{D}|$. Given this, we must find $\mathcal{T}_{ML}(\mathcal{D})$ by generating multiple samples and aggregating worker responses across them.

To find $\mathcal{T}_{ML}(\mathcal{D})$, \textsc{Orchestra} has an iterative refinement procedure that performs repeated iterations of (\textsc{GenerateSample} $\to$ \textsc{ConstructHierarchy} $\to$ \textsc{MergeHierarchies} $\to \dots$), to generate a final hierarchy. At the end of each iteration, we generate a new estimate for $\mathcal{T}_{ML}(\mathcal{D})$. We give an intuitive explanation for these algorithms below; a detailed description is given in the next section. 
\begin{denselist}
\item {\sc GenerateSample.} Any sample of items that we generate must contain some item overlap with previously generated samples, as well as contain new items that explore the dataset. The overlap helps us locate worker frontiers on this sample within the current estimate of $\mathcal{T}_{ML}(\mathcal{D})$, while the new items allow us to expand 
$\mathcal{T}_{ML}(\mathcal{D})$ by finding new concepts. We provide a procedure to check if two workers---working on different samples---are providing frontiers on the same latent hierarchy.
\item {\sc HierarchyConstruction.} The construction algorithm takes as input multiple worker frontiers collected for a single sample, and outputs the dominant hierarchy. To separate the dominant hierarchy, \textsc{HierarchyConstruction} infers whether these 
 frontiers are chosen from the same hierarchy, or different ones.
\item {\sc MergingHierarchies.} To combine hierarchies across multiple samples, the merging algorithm takes as input two hierarchies --- the current estimate of $\mathcal{T}_{ML}(\mathcal{D})$, and the hierarchy constructed on the current sample. The output is a new estimate of $\mathcal{T}_{ML}(\mathcal{D})$, and is calculated by augmenting the current estimate of $\mathcal{T}_{ML}(\mathcal{D})$. The merging exploits the location of the overlap items in the current estimate of $\mathcal{T}_{ML}(\mathcal{D})$.
\end{denselist}
At the end of this iterative procedure, we return the maximum likelihood frontier in $\mathcal{T}_{ML}(\mathcal{D})$ as the consensus clustering. The quality of the consensus clustering depends on whether the number of iterations were sufficient to ensure that most items in $\mathcal{D}$ can be categorized into this consensus clustering. 
In the next section we provide the details of the workflow for \textsc{Orchestra} --- including a sampling guarantee that gives a lower bound on the size of the samples needed to cover \emph{atleast} some fraction of items in $\mathcal{D}$.

%% file: creating_hierarchies.tex
\section{Orchestra Workflow}
\label{sec:3}
As noted in the previous section, workers may choose different frontiers in different latent hierarchies when asked to cluster a set of items. The problem of finding the maximum likelihood hierarchy is then equivalent to finding a hierarchy that {\it best explains} the most worker clusterings. In this section, we provide algorithms to find this hierarchy, as well as the consensus clustering within that hierarchy. We also give theoretical results that allow us to limit the number of iterations in the \orc workflow. First, in Section~\ref{sec:hierarchyConstruc}, we describe the {\sc HierarchyConstruction} algorithm that finds the most likely hierarchy under the assumption that all workers cluster the same set of $n$ items. Then, in Section~\ref{sec:Sampling Guarantee}, we provide a guarantee that helps us fix a reasonable value for $n$. Section~\ref{sec:generateSample} lays out the {\sc GenerateSample} algorithm, and in Section~\ref{sec:mergingHierarchies}, we generalize our setting with the {\sc MergingHierarchies} algorithm, allowing workers to cluster different subsets of items, and aggregating their clusterings to get a single hierarchy. Finally, in Section~\ref{sec:categorize}, we present a procedure to find the consensus clustering from the final hierarchy that our iterative workflow generates, as well as describing how we categorize items. 

\papertext{Due to space limitations, we
omit all proofs and pseudocode; they can be found in our extended
technical report~\cite{orchestra2015}.}

\begin{figure*}[t!]
\subfigure[Examples of real worker clusterings for the dataset in Figure~\ref{fig:clusterinterface}.]{\includegraphics[width=0.48\linewidth]{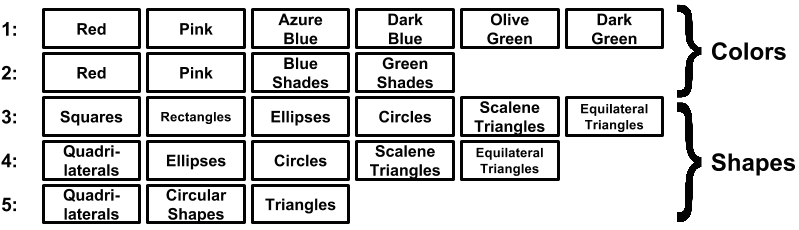}\label{fig:figA}}
\subfigure[The hierarchy $\mathcal{T}$ corresponding to the maximum sized clique $3,4,5$ in (c) using {\sc ConstructHierarchy}.]{\includegraphics[width=0.42\linewidth]{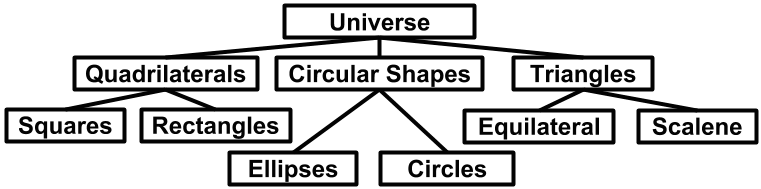}\label{fig:figC}}
\subfigure[The clustering graph for the worker clusterings shown in (a).]{\includegraphics[width=0.18\linewidth]{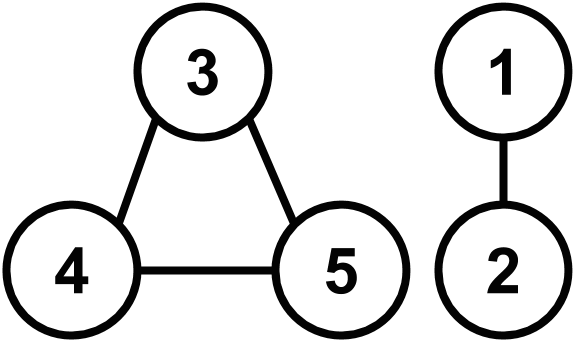}\label{fig:figB}}\qquad
\subfigure[A hypothetical hierarchy $\mathcal{T}(\mathcal{S})$ constructed in the 2nd iteration of our workflow, which contains an extra {\tt Hexagons} concept.]{\includegraphics[width=0.33\linewidth]{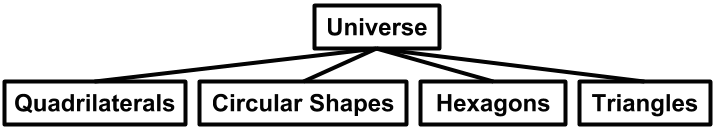}\label{fig:figD}}\qquad
\subfigure[The hierarchy $\mathcal{T}'$ constructed by merging (b) and (d) using {\sc MergingHierarchies} after 2 iterations.]{\includegraphics[width=0.42\linewidth]{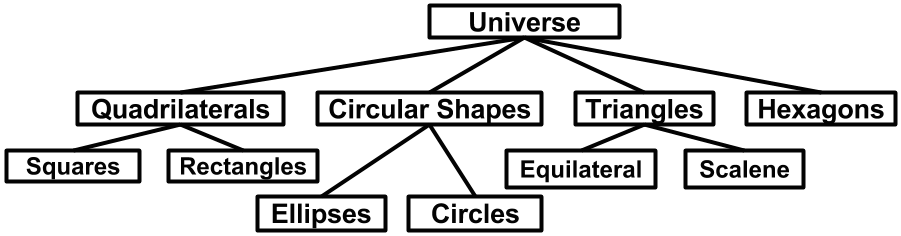}\label{fig:figE}}
\vspace{-10pt}
\caption{An example demonstrating our iterative workflow approach on the Shapes dataset of Figure~\ref{fig:shapes}.}
\label{fig:workflowexample}
\vspace{-10pt}
\end{figure*}

\subsection{\label{sec:hierarchyConstruc}The \textbf{\sc \large HierarchyConstruction} Algorithm}
Given a set of items $\mathcal{S} = \{x_{1},\dots, x_{n}\} \subseteq \mathcal{D}$, we ask $m$ workers to cluster the items in $\mathcal{S}$. We denote the set of worker clusterings by $\mathfrak{C} = \{\mathbb{C}_1, \dots, \mathbb{C}_m\}$, where $\mathbb{C}_i = \{C_{i, 1}, ..., C_{i, k_i}\}$ is the set of clusters proposed by worker $i$. Note that workers can give as many clusters as they like, but no cluster is allowed to be empty. Figure~\ref{fig:figA} shows some clusterings proposed by workers on the sample of items shown in Figure~\ref{fig:clusterinterface}.


\vspace{-2pt}
\begin{problem}[\textbf{Hierarchy Construction}]\label{prob:hierarchy}
Given the clusterings $\mathfrak{C}$ on a set of items $\mathcal{S}$, find a hierarchy $\mathcal{T}(\mathcal{S})$ such that the number of clusterings from $\mathfrak{C}$ that can be associated with complete frontiers in $\mathcal{T}(\mathcal{S})$ is maximum.  
\end{problem}
\vspace{-4pt}
\noindent Intuitively, we would like to find the maximum likelihood hierarchy, \emph{i.e.}, one that contains the maximum number of clusterings as complete frontiers. For instance,  clustering $5$ in Figure~\ref{fig:figA} can be associated with Figure~\ref{fig:figC} as a complete frontier, covering all items in the dataset.

We will show that Problem~\ref{prob:hierarchy} is equivalent to the {\sc Max-Clique} problem. {\sc Max-Clique} refers to the problem of finding the maximum sized clique in a graph $G$, and is a well-known {\sc np-hard} problem. Consequently, the optimal solution takes exponential time to compute. However, in our case, the graph for which {\sc Max-Clique} must be solved is small, so the computation is still feasible. We will prove the equivalence to {\sc Max-Clique} via a constructive proof. We first provide some definitions that will help us carry out the construction.


\begin{definition}[\textbf{Consistency of Clusterings}]
\label{def:Consistency}
Clusterings $\mathbb{C}_i = \{C_{i,1}, \dots, C_{i, k_i}\}$ and $\mathbb{C}_j = \{C_{j, 1}, \dots, C_{j, k_j}\}$ are said to be consistent if and only if for every $(s,t) \in \{1, \dots, k_i\} \times \{1, \dots, k_j\}$, one of the following holds:
\vspace{-7pt}
\begin{multicols}{2}
\begin{enumerate}
\item[(1)] $C_{i, s} \cap C_{j, t} = \phi$
\item[(2)] $C_{i, s} \subset C_{j, t}$
\item[(3)] $C_{i, s} \supset C_{j, t}$
\item[(4)] $C_{i, s} = C_{j, t}$
\end{enumerate}
\end{multicols}
\end{definition}
\vspace{-7pt}
\noindent 
In Figure~\ref{fig:figA}, the worker clusterings $1$ \& $2$ are consistent --- {\tt Blue Shades} decomposes perfectly into {\tt Azure Blue} and {\tt Dark Blue}, as does {\tt Green Shades} --- while $1$ is inconsistent with $3,4,5$.
Since every clustering is associated with a frontier, we can also define a corresponding notion of \emph{consistent frontiers}: we simply replace $\supset$ by $\succ$ in Definition~\ref{def:Consistency}. It is useful to note that any two complete frontiers in the same hierarchy will always be consistent.  In Figure~\ref{fig:Hier1}, the complete frontiers \{{\tt Quadrilaterals}, {\tt Triangles}, {\tt Ellipses}, {\tt Circles}\} and \{{\tt Squares}, {\tt Rectangles}, {\tt Triangles}, {\tt Round}\} are consistent. 

\vspace{-2pt}
\begin{definition}[\textbf{Clustering Graph}]
Clustering graph $G_{\mathfrak{C}} = (\mathfrak{C}, E)$ is an undirected graph, where each clustering in $\mathfrak{C}$ corresponds to a unique vertex in $G$ and there is an edge between $\mathbb{C}_i$ and $\mathbb{C}_j$ $\forall \, i,j \in \{1,\dots,m\}$ if and only if $\mathbb{C}_i$ and $\mathbb{C}_j$ are consistent.
\end{definition}
\vspace{-2pt}

\noindent Figure~\ref{fig:figC} depicts the clustering graph for the clusterings shown in Figure~\ref{fig:figA}. Each worker clustering corresponds to a node in the graph. Notice how there is no edge from $1$ to any of $3,4,5$, since they are mutually inconsistent.

Let $\mathfrak{C}_{\mathrm{CLIQUE}} \subseteq \mathfrak{C}$ be a clique in $G_{\mathfrak{C}}$. Let the set of all \emph{unique} clusters in $\mathfrak{C}_{\mathrm{CLIQUE}}$ be $\mathcal{H} = \{C_{i, j} \mid C_{i, j} \in \mathbb{C}_i, \forall \, \mathbb{C}_i \in \mathfrak{C}_{\mathrm{CLIQUE}}\}$. $\mathcal{H}$ can be organized into a hierarchy $\mathcal{T}_{\mathcal{H}}$ as follows: for every cluster $ C_{i, j} \ \in \ \mathcal{H}$, find the smallest cluster in $\mathcal{H} \cup \text{{\tt Universe}}$ that is a superset of $ C_{i, j}$ and mark that as the parent of $ C_{i, j}$ in $\mathcal{T}_{\mathcal{H}}$. 
\techreport{Algorithm~\ref{alg1} shows the pseudocode for this {\sc HierarchyConstruction} algorithm.} 

Consider the clique $3,4,5$ in the clustering graph of Figure~\ref{fig:figC}. $\mathcal{H}$ contains a total of 14 clusters as shown in Figure~\ref{fig:figA}. Suppose we wanted to find the parent of {\tt Rectangles}; the smallest cluster in $\mathcal{H} \ \cup$ {\tt Universe} containing {\tt Rectangles} is {\tt Quadrilaterals}. The cluster {\tt Universe} also contains {\tt Rectangles} but it is not the smallest such cluster. Thus, we make {\tt Quadrilaterals} the parent of {\tt Rectangles}, as shown in Figure~\ref{fig:figB}. Similarly, {\tt Universe} becomes the parent of {\tt Quadrilaterals}. The hierarchy after this construction is shown in Figure~\ref{fig:figB}.

We state the following lemma and theorem which show that our construction is valid, and omit the proof. As mentioned earlier, all proofs can be found
in our extended technical report~\cite{orchestra2015}.
\vspace{-4pt}
\begin{lemma}
\label{lemma:first}
For any $ C_{i, j} \in \mathcal{H}$, the smallest cluster in $\mathcal{H} \, \cup \, \text{{\tt Universe}}$ that is a superset of $ C_{i, j}$, is unique.
\vspace{-5pt}
\end{lemma}
\techreport{
\begin{proof}
Since {\tt Universe} is the superset of all clusters in $\mathcal{H}$, every $ C_{i, j}$ has at least one superset in $\mathcal{H} \cup \ \text{{\tt Universe}}$. Assume that there are two distinct smallest clusters ${C}_{1,x}$ and $C_{2,y}$ that are both supersets of $ C_{i, j}$. It follows that the clusterings to which ${C}_{1,x}$ and $C_{2,y}$ belong \emph{i.e.} $\mathbb{C}_{1}$ and $\mathbb{C}_{2}$ cannot be consistent. This can be seen by noting that ${C}_{1,x}$ and $C_{2,y}$ do not satisfy any of the four conditions of Definition~\ref{def:Consistency}. This contradicts the fact that $\mathbb{C}_{1}$ and $\mathbb{C}_{2}$ are part of the same clique in the clustering graph, and the result follows.
\end{proof}
}
\begin{theorem}
$\mathcal{T}_{\mathcal{H}}$ is a hierarchy.
\end{theorem}

Suppose we pick $\mathfrak{C}_{\mathrm{CLIQUE}}$ to be the maximum sized clique in $G_{\mathfrak{C}}$, and let $\mathcal{T}_{\mathrm{max}}$ be the hierarchy that is generated using this clique. We now state an important result.
\begin{theorem}
Suppose every clustering $\mathbb{C} \in \mathfrak{C}$ lies in exactly one maximal clique.
Also suppose the total number of latent hierarchies is $k$. Then, $\mathcal{T}_{\mathrm{max}}$ is the maximum-likelihood hierarchy with probability atleast $\left[1 - \left(1 - \frac{1}{k}\right)^m\right]$, where $m$ is the number of workers. 
\end{theorem}
\noindent Figure~\ref{fig:figB} shows the maximum likelihood hierarchy corresponding to the maximal clique $3,4,5$ in the clustering graph of Figure~\ref{fig:figC}. Notice that this hierarchy corresponds to organizing the dataset on {\sc shape}. 

Identifying the most probable hierarchy is thus equivalent to finding the maximum sized clique in $G_{\mathfrak{C}}$. The size of $G_{\mathfrak{C}}$ is atmost the number of workers $m$, since we can combine identical worker clusterings into a single node. Since $m$ is typically small $(\le 15)$, solving {\sc Max-Clique} is quite tractable.

We note that we do not provide an explicit mechanism for worker mistakes. In our experiments, we find that workers made no errors with respect to the organization they had in mind. Even if some workers do make errors, our maximum likelihood hierarchy only contains those workers who clustered items consistently, and so either these errors would not be incorporated, or a large number of workers would have to make these errors in the same way, which is unlikely. In the future, we plan to relax our definition of consistency of clusterings, to admit a small tolerance threshold.

\input{sampling.tex}

\input{generate_sample.tex}

\input{merging.tex}

\input{categorization.tex}

%% file: sampling.tex
\subsection{\label{sec:Sampling Guarantee}Sampling Guarantee}

As discussed in Section~\ref{sec:prelim}, our approach is to first construct a maximum likelihood hierarchy using several samples and then categorize the remaining items into the maximum likelihood frontier in this hierarchy. Recall that for $\mathcal{S} \subseteq \mathcal{D}$, complete frontiers in a hierarchy $\mathcal{T}(\mathcal{S})$ are not generally complete in $\mathcal{T}(\mathcal{D})$. This is because some concepts may have instances in $\mathcal{D}$, but not in $\mathcal{S}$. For instance, suppose $\mathcal{D}$ contains instances of {\tt Ellipses} and {\tt Circles} while $\mathcal{S}$ only contains items from {\tt Ellipses}, but no instance of {\tt Circles}. Then, $\mathcal{T}(\mathcal{S})$ would not contain the {\tt Circles} concept. {\tt Circles} would therefore go {\it undiscovered} in our sample.
We now prove a guarantee that allows us to make a suitable choice for the parameter $n; |\mathcal{S}| = n$.

Intuitively, if the size of $\mathcal{S}$ is large, we can be confident that the sample will {\it discover} the concepts that occur frequently in the dataset, \emph{i.e.,} the concepts that have many instances in the dataset. On the contrary, when $\mathcal{S}$ is small, a large number of concepts are likely to go undiscovered. Thus the hierarchy constructed on a small sample may not be representative of the entire dataset. We formalize the notion of {\it representativeness} below.

\begin{definition}[\textbf{Concept Coverage}]
The coverage of a concept $C$ is the fraction of items in $\mathcal{D}$ that are instances of $C$.
\end{definition}

\noindent We say that a sample $\mathcal{S}$ \textit{discovers} $C$ if and only if there is an item $s \in \mathcal{S}$ that is an instance of $C$.

\begin{definition}[\textbf{Frontier Coverage}]
\label{define:frontier_coverage}
The coverage of a frontier $F$ with respect to a sample $\mathcal{S}$ is the sum of coverages of the concepts in $F$ that $\mathcal{S}$ \textit{discovers}.
\end{definition}
Suppose that $\mathcal{S}$ contains $n$ randomly sampled items. Let $\mathcal{T}(\mathcal{D})$ be a hierarchy constructed on $\mathcal{D}$ and let $F$ be a complete frontier in $\mathcal{T}(\mathcal{D})$ containing $f$ concepts. We give a lower bound on the expected coverage of $F$ with respect to $\mathcal{S}$, $\mathbb{E}_{\mathcal{S}}[X_F]$ below. 

\begin{theorem}
\label{theorem:SamplingGuarantee}
$\mathbb{E}_{\mathcal{S}}[X_F] \geq 1 - \frac{f}{n + 1}(1 - \frac{1}{n + 1})^n$
\end{theorem}
\techreport{
\begin{proof}
Let $p_i$ be the coverage of the $i^{th}$ concept in $F$. Let $X_{F, i}$ be a random variable that equals 0 if $\mathcal{S}$ does not discover the $i^{th}$ concept of $F$ and equals $p_i$ otherwise. Using Definition~\ref{define:frontier_coverage}, the coverage of $F$ is $X_F = \displaystyle \sum_{i = 1}^{f} X_{F, i}$. We have,
\begin{align*}
\mathbb{E}[X_F] &= \displaystyle \sum_{i = 1}^{f}\mathbb{E}[X_{F, i}]=\displaystyle \sum_{i = 1}^{f}p_i(1 - (1 - p_i)^n)\\
&= \displaystyle 1 - \sum_{i = 1}^{f}p_i(1 - p_i)^n \geq \displaystyle 1 - \max_{\substack{0\leq p_{i}\leq 1\\ \sum_{i}p_{i} = 1}}\sum_{i = 1}^{f}p_i(1 - p_i)^n \\
&\geq \displaystyle 1 - \max_{0\leq p_{i}\leq 1}\sum_{i = 1}^{f}p_i(1 - p_i)^n
\end{align*}
Now, for $p_{i} \ \in \ [0, 1]$, $p_i(1 - p_i)^n$ is maximum when $p_i = \frac{1}{n + 1}$ and the result follows.
\end{proof}
}

\begin{figure}[t!]
\centering
\hspace{-10pt}
\subfigure{\includegraphics[scale=0.16]{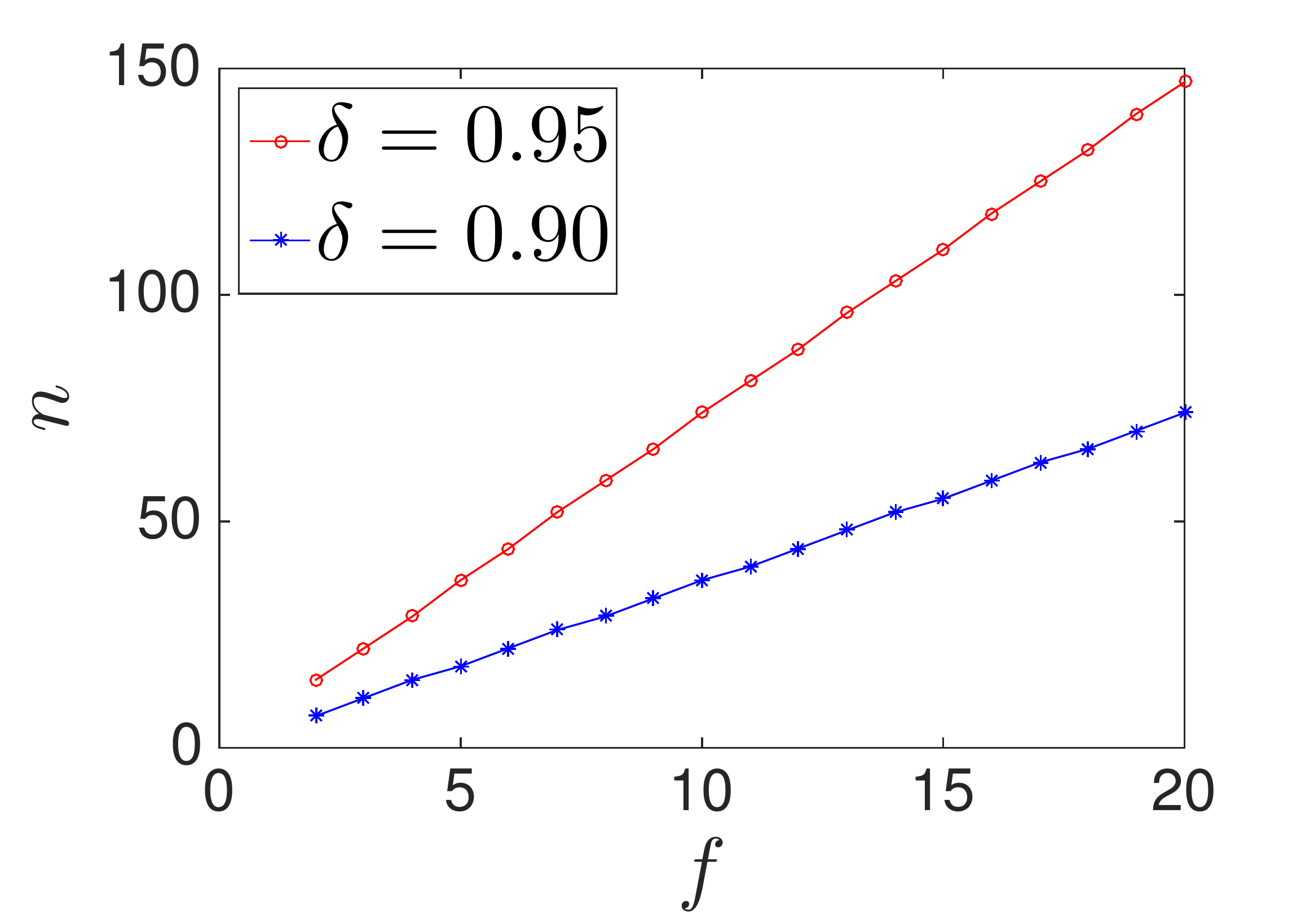}\label{fig:samplingGuarantee}}
\hspace{-10pt}
\subfigure{\includegraphics[scale=0.16]{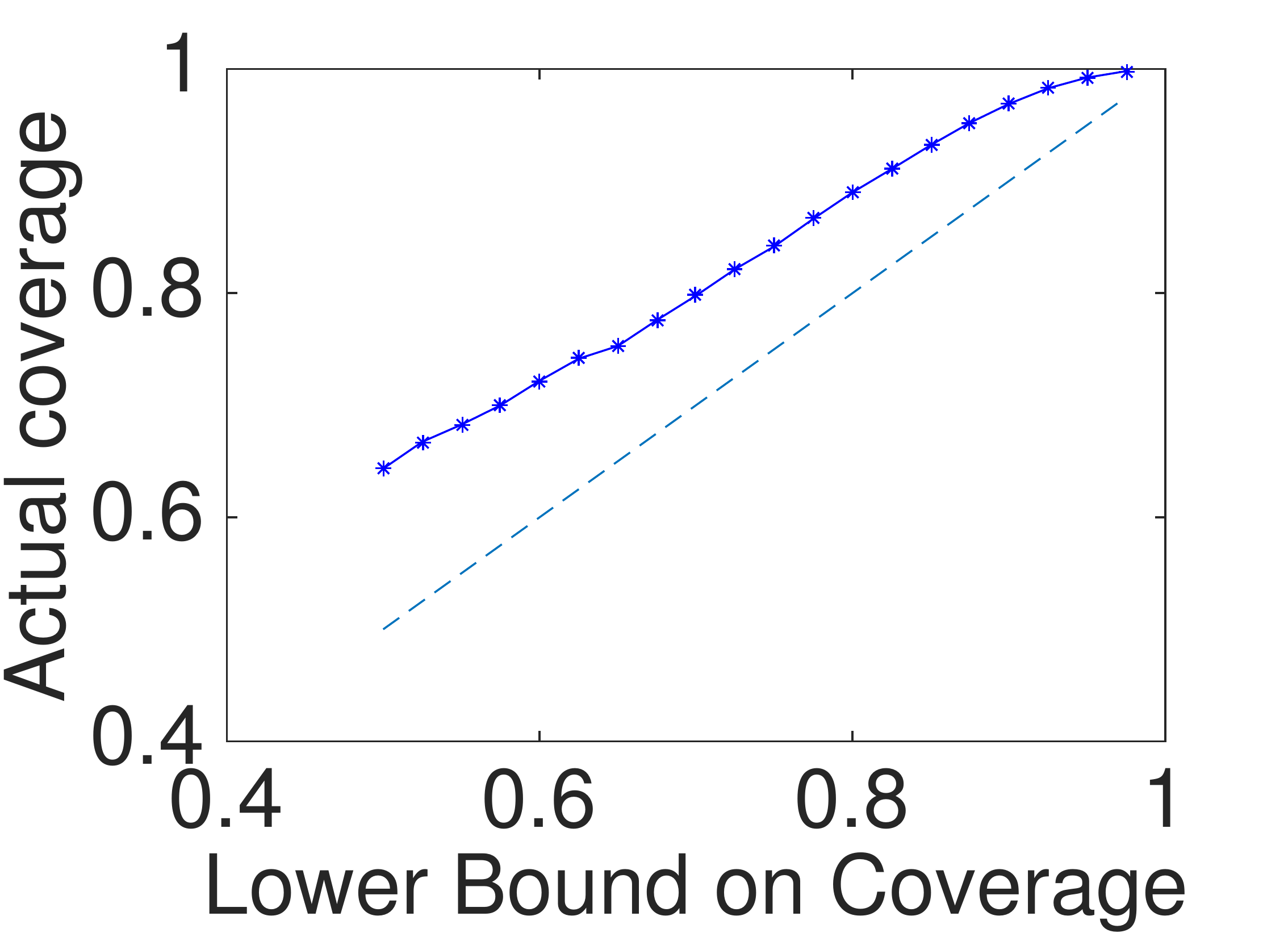}\label{fig:samplingGuarantee2}}
\vspace{-10pt}
\caption{(a) Plot showing sample size $n$ vs. size of complete frontier $f$ to ensure $\delta$ expected coverage. (b) Plot showing lower bound on expected coverage vs. actual expected coverage over 1000 trials -- the dotted line is the $45\deg$ line.}
\vspace{-13pt}
\end{figure}

\noindent In Figure~\ref{fig:samplingGuarantee}, we plot $n$ vs. $f$ for different values of a threshold $\delta$, which lower bounds the expected coverage. Observe that for fixed values of $\delta$, $n$ increases linearly with $f$. However, this lower bound is not tight and the actual coverage turns out to be greater than $\delta$. To observe this, we plot $\delta$ vs. the actual coverages that we get in Figure~\ref{fig:samplingGuarantee2}. For each value of $\delta$, we pick 20 $(f,n)$ pairs that satisfy the bound and conduct 1000 random trials for each pair. Each trial consists of assigning a random probability distribution to $f$ bins, which gives the probability of an item in the dataset belonging to a bin (concept). We then draw $n$ samples from this distribution, and compute the actual coverage that we get. 


We can similarly find an upper bound for the variance of $X_F$.
\begin{theorem}
$\mathrm{Var}_{\mathcal{S}}[X_F] \le 1 - \left(1 - \frac{f}{n+1}\left(1 - \frac{1}{n+1}\right)^{n} \right)^2$
\end{theorem}
\noindent As a rule of thumb, we pick $\delta = 0.95$ and $f = 16$ for the experiments that we carry out, which results in $n = 115$.
For $n = 115$ and $f = 16$, the variance is upper bounded by $0.1$.
However, we note that we cannot expect a single worker to be able to organize
$115$ items at once, so we describe how to iteratively cluster smaller sets
of items while being able to combine the information across these iterations
in the next section. 

%% file: generate_sample.tex
\subsection{\label{sec:generateSample}The {\sc \large GenerateSample} Algorithm}
As we noted earlier, it is not possible for a single worker to generate a clustering for large $\mathcal{S}$, especially one as large as $\approx 120$. Instead, our approach will be to repeatedly instantiate smaller $\mathcal{S}$ for every iteration, while bounding its size. 
For example, one approach would be to instantiate four distinct sets $\mathcal{S}$ of size $30$ each,
each of which is organized by workers. 
However, due to the lack of overlap across these sets, it is impossible to relate
the hierarchies constructed across these sets to one another.
Intuitively, for each new iteration, it is desirable that $\mathcal{S}$ contains some item overlap with the current estimate of the maximum likelihood hierarchy (at the end of the previous iteration), so that the new hierarchy we generate can be easily merged into the current estimate of the maximum likelihood hierarchy. We now provide a mechanism to fix the size of this overlapping set, which we call the \emph{kernel} of $\mathcal{S}$. Each sample $\mathcal{S}$ consists of some new items, not encountered before, along with items that have already been organized in the current maximum likelihood hierarchy, which constitute $K$, the kernel of $\mathcal{S}$.


To set a reasonable value for the kernel $K$, we need it to be large enough to allow us to perform hierarchy merging at each iteration. It also needs to be small enough to allow introduction of some new items into our sample. Our strategy for picking the kernel items is to pick a single item from each of the leaf nodes in the current hierarchy estimate. Therefore, we set $|K| = $ \# of leaves and randomly sample the rest of the items in $\mathcal{S}$ from $\mathcal{D}$. 

The justification for this strategy is that sampling a single item from every leaf allows us to determine any concept a worker generates --- whether an internal node in the hierarchy, or a leaf in our 
current maximum likelihood hierarchy. If a worker combines some kernel items into a single cluster, we can infer the concept of the entire cluster (which includes some new items) by finding where these kernel items occur together in our hierarchy. We will make this idea more precise in the {\sc MergingHierarchies} algorithm. 

In our experiments, we have never encountered a case 
where $|K|$ is too large:
nevertheless, in such cases,
we can simply split $|K|$ up into equal sized smaller 
portions, and repeat the clustering of the same set 
of new items with these smaller portions of the kernel,
such that each of the new items gets the opportunity
to be associated with or clustered with any of the kernel items. 

%% file: merging.tex
\subsection{\label{sec:mergingHierarchies}The {\sc \large MergingHierarchies} Algorithm}
In this subsection, we describe our {\sc MergingHierarchies} algorithm, in which we make use of the kernel formulation that we introduced above. 

Let $\mathcal{T}$ be our current estimate of the maximum likelihood hierarchy generated after the $\tau^{th}$ iteration. Suppose that we run {\sc HierarchyConstruction} on the sample $\mathcal{S}$ generated for the $(\tau + 1)^{th}$ iteration, and get a hierarchy $\mathcal{T}(\mathcal{S})$. Using $K$, the kernel of $\mathcal{S}$, we would like to merge $\mathcal{T}(\mathcal{S})$ into $\mathcal{T}$ to generate a new hierarchy $\mathcal{T}'$, by mapping known concepts across these hierarchies. To carry out this merging, we will assume that the kernel items in a cluster represent that cluster's concept accurately, as well as any super-concepts (i.e., concepts that are ancestors of the cluster's concept).

Consider the set of leaf nodes $C_1,\dots,C_l$ in $\mathcal{T}(\mathcal{S})$ and let the set of kernel items in $C_i$ be $K_i$. For each leaf $C_i$: 

\noindent {\small $\bullet$} Suppose $|K_i| > 0$ for $C_i$; we map $C_i$ to the node $C$ in $\mathcal{T}$ that contains the smallest superset of $K_i$. This is simply the lowest common ancestor of the leaf nodes in $\mathcal{T}$ that contain kernel items from $K_i$. Intuitively, if the kernel items are identical then both clusters are associated with the same concept, and can therefore be merged. All items in $C_i$ are transferred to $C$.

\noindent {\small $\bullet$} Suppose $|K_i| = 0$ for $C_i$; we first find the ancestor $C_a$ (with kernel $K_a$) closest to $C_i$ (the lowest ancestor) in $\mathcal{T}(\mathcal{S})$ such that $|K_a| > 0$. Since $C_i$ contained no kernel items, it is clear that $C_i$ represents some new concept; we must search for another concept that {\em generalizes} $C_i$. As before, we map $C_a$ to the node $C$ in $\mathcal{T}$ that contains the smallest superset of $K_a$. However, since we need to map $C_i$ and not $C_a$, we instead insert $C_i$ as a new child of $C$ in $\mathcal{T}$.

After mapping all leaf nodes in $\mathcal{T}(\mathcal{S})$ to $\mathcal{T}$, we get a new maximum likelihood hierarchy $\mathcal{T}'$ after the $(\tau+1)$th iteration. It is easy to see that $\mathcal{T}'$ is indeed a hierarchy. The only changes we make are \emph{(a)} adding in new items to the concept of which they are instances, which does not modify the hierarchy and \emph{(b)} adding in new concept nodes. For \emph{(b)}, notice that by construction, we attach the new concept $C_i$ to the lowest concept $C$ that generalizes it. $C_i$ is disjoint with respect to all other children of $C$, otherwise it would contain a kernel item. $C_i$ is also necessary to allow all items to exist at the leaf nodes, since no other child of $C$ covers the concept discovered in $C_i$. Therefore, $\mathcal{T'}$ is a hierarchy.

Figure~\ref{fig:workflowexample} demonstrates an example of {\sc MergingHierarchies}. Figure~\ref{fig:figC} is our current estimate $\mathcal{T}$ to be merged with Figure~\ref{fig:figD}, depicting $\mathcal{T}(\mathcal{S})$. The merged hierarchy $\mathcal{T}'$ is shown in Figure~\ref{fig:figE}. By our {\sc GenerateSample} algorithm, the kernel of $\mathcal{S}$ would contain 6 items, one each for the leaves of $\mathcal{T}$ in Figure~\ref{fig:figC}. Even though $\mathcal{T}(\mathcal{S})$ combines the kernel items corresponding to {\tt Squares} and {\tt Rectangles} into the {\tt Quadrilaterals} cluster in Figure~\ref{fig:figE}, we can map {\tt Quadrilaterals} in $\mathcal{T}(\mathcal{S})$ using these 2 kernel items, to the {\tt Quadrilaterals} cluster in $\mathcal{T}$, the lowest node where they occur together. For the {\tt Hexagons} cluster in $\mathcal{T}(\mathcal{S})$, we first find its deepest ancestor in $\mathcal{T}(\mathcal{S})$ that contains a kernel item, which turns out to be {\tt Universe}. {\tt Universe} in $\mathcal{T}(\mathcal{S})$ is mapped to {\tt Universe} in $\mathcal{T}$, and {\tt Hexagons} is inserted as a child, as shown in Figure~\ref{fig:figE}.

\techreport{
\begin{algorithm}                      
\caption{\texttt{MergingHierarchies}($\mathcal{T},\mathcal{T}(\mathcal{S}))$}          
\label{alg3}                           
\begin{algorithmic}                    
    \REQUIRE current hierarchy estimate $\mathcal{T,}$ generated hierarchy $\mathcal{T}(\mathcal{S})$
    \ENSURE Hierarchy $\mathcal{T}'$
        \FOR{each leaf node $C_i \in \mathcal{T}(\mathcal{S})$}
    \STATE $K_i \leftarrow$ kernel items in $C_i$
    \STATE $F \leftarrow \{\}$
    	\IF{$|K_i|$ > 0}
		\FOR{$x \in K_i$}
			\STATE $F \leftarrow F \ \cup$ leaf node in $\mathcal{T}$ containing $x$
		\ENDFOR
		\STATE $C \leftarrow $ deepest common ancestor of $F$
		\STATE $C \leftarrow C \cup C_i$
    	\ELSE
		\STATE $C_a \leftarrow$ deepest ancestor of $C_i$ with some kernel item(s) $K_a$
		\FOR{$x \in K_a$}
			\STATE $F \leftarrow F \ \cup$ leaf node in $\mathcal{T}$ containing $x$
		\ENDFOR
		\STATE $C \leftarrow $ deepest common ancestor of $F$
		\STATE Parent($C_i$) $\leftarrow C$
    	\ENDIF
    \ENDFOR
    
\end{algorithmic}
\end{algorithm}
}
We now conclude this section with a discussion of the cost of our iterative workflow.

\stitle{Cost of Iterative Workflow.}
Our sampling guarantee requires us to sample $n$ items, while the kernel overlap is fixed to be the \# of leaves in the current hierarchy estimate. Notice that when fixing the value of $n$, we assumed that the size of a complete frontier in the dataset hierarchy is $f$. We do not expect our choice of $n$ to discover \emph{any more than} an $f$-sized complete frontier. We can therefore upper-bound the value of $|K|$, the size of the kernel, to be $f$, since we would not expect the number of leaves in our maximum likelihood hierarchy to exceed $f$. Assume that in each iteration, we ask for clusterings on $h$ items. To find the total number of iterations $\tau$, we find the smallest value that satisfies
$
h + (h - f)(\tau - 1) \ge n
$, where we have replaced $|K|$ with its upper bound $f$ in each iteration. 
We typically set $h = 35$. For $f = 16$ and $n = 115$, $\tau$ turns out to be 6. If each iteration is clustered by $m$ workers, the total cost of our iterative workflow becomes $\mathrm{O}\left(m\left\lceil \frac{(n - f)}{(h - f)}\right\rceil\right)$. The cost of our workflow is \emph{independent of the size of the dataset} $\mathcal{D}$.

\techreport{\agp{Maybe this is for later, but feels like a table of notation would be valuable, along with rules of thumb. Ideally we should we playing around with these parameters in our experiments too....}}

%% file: categorization.tex
\subsection{\label{sec:categorize}Categorization}
At the end of our iterative workflow, we have a final maximum likelihood hierarchy $\mathcal{T}$, from which we must extract the consensus clustering granularity or frontier. As we stated in Section~\ref{sec:prelim}, the reason we construct this hierarchy is to preserve information about the granularities at which workers cluster items. We can now simply determine the consensus clustering \emph{i.e.} the granularity workers are most likely to cluster on. This consensus clustering is the maximum-likelihood complete frontier in $\mathcal{T}$. We now outline a procedure to find this frontier. Subsequently we describe how we use this frontier for categorization.

\stitle{Maximum-Likelihood Frontier.} Suppose that $\mathcal{T}$ consists of the set of nodes $V$ with root node $R$. Associate with each node $v$, an event $E_v$ that $v$ is split by a worker \emph{i.e.,} a worker chooses to give us nodes/concepts below $v$ in their frontier. Let $F$ be a frontier in $\mathcal{T}$ and let $A$ be the set of ancestors of $F$, excluding $R$. We define the likelihood of $F$ as,
\begin{multline*}
\mathcal{L}(F) = p(E_R)\prod_{v \in F} p(\overline{E_v} | E_{v(1)},\dots, E_{R}) \\ \prod_{v' \in A} p(E_{v'} | E_{v'(1)},\dots, E_{R})
\end{multline*}
where $v(1), \dots,R$ are the ancestors of $v$. $v(1)$ is the parent of $v$, $v(2)$ is the parent of $v(1)$, etc. Observe that $p(E_v | E_{v(1)}, \dots, $$E_{v(k)}) = p(E_v | E_{v(1)})$ \emph{i.e.,} $E_v$ is conditionally independent of the rest of its ancestors, given its immediate parent. We have,
\begin{equation*}
\mathcal{L}(F) = p(E_R)\prod_{v \in F} p(\overline{E_v} | E_{v(1)}) \prod_{v \in A} p(E_v | E_{v(1)}) \\
\end{equation*}
Given a set of worker responses (complete frontiers), we approximate $p(\overline{E_v} | E_{v(1)})$ as the ratio of the number of workers who gave node $v$ as a frontier to the total number of workers who gave node $v$ or its descendants as frontiers. Note that $p(E_R) = 1$. We are interested in $\underset{F}{\mathrm{argmax}}\ \mathcal{L}(F)$ which can be computed using the recurrence relation below.
\begin{equation*}
D(v) = \mathrm{max}\left(p(\overline{E_v} | E_{v(1)}), p(E_v | E_{v(1)}) \prod_{u \in C(v)} D(u)\right)
\end{equation*}
where $C(v)$ is the set of children of $v$. Intuitively, at node $v$, there are two choices: either keep node $v$ as the frontier, or drill down and check for more likely frontiers, and using this, we can find the maximum likelihood frontier.

\stitle{Categorization on Frontier.} To carry out categorization we present workers with an interface similar to the clustering interface of Figure~\ref{fig:clusterinterface} with certain key differences. For each cluster in our consensus maximum likelihood clustering, we give a few `pivot' items to the worker as exemplars for that cluster. Our assumption is that these pivots completely capture the concept represented by that cluster. Workers are asked to categorize items into the cluster which seems most appropriate. Once again, we do not point workers to any attributes in the data, instead relying on our pivots to allow them to infer the organization of our consensus clustering. 

The cost of our categorization step is easily calculated --- there are $|\mathcal{D}| - n$ items left after the clustering phase, and suppose we take $\theta$ votes per item. Typically $n \ll |\mathcal{D}|$, so the total categorization cost is then $\mathrm{O}( \theta|\mathcal{D}|)$, linear in the size of the dataset. 

%% file: experiments.tex
\label{sec:exp}
In our experiments, our goal is to qualitatively 
and quantitatively compare \orc to other 
crowd-powered clustering algorithms.

\stitle{Datasets.} We used the following datasets
in our experiments. 
\begin{denselist}
\item 
Our first dataset, titled {\em shapes}, 
is a synthetic, stylized dataset
consisting of shapes, our running
example in Section~\ref{sec:prelim}.
As described, each item is a random 
({\sc shape}, {\sc size}, {\sc color}) tuple.
The images from this dataset are also the ones
displayed in Figure~\ref{fig:clusterinterface}.
We use this dataset because we can control 
the organizational hierarchies in this dataset,
allowing us to evaluate the performance of
algorithms on recovering clusterings across
one or more hierarchies in the dataset.
\item 
The second dataset, titled {\em scenes}, contains images
from 13 categories~\cite{fei2005bayesian}
in natural and man-made surroundings.
This dataset was also used in prior work on
crowd clustering~\cite{gomes2011crowdclustering,yi2012crowdclustering}.
\item 
The third dataset, titled {\em imagenet},
contains images from ImageNet\cite{deng2009imagenet}. 
Images are sampled randomly from 20 categories:
\{buildings, cars, parrots, vulture, fruit, flower, vegetable, fighter, commercial, helicopter, ship, seahorse, whale, cheetah, lion, elephant, tiger, jellyfish, sparrow, leaves\}.
\end{denselist}
Across all datasets, we conduct multiple runs 
of different algorithms on random subsets of the datasets.

\stitle{Algorithms.}
We compare the following state-of-the-art crowd-powered
clustering algorithms with \orc:
\begin{denselist}
\item \texttt{CrowdClust}: This is the algorithm from Gomes et al.~\cite{gomes2011crowdclustering}. 
\item \texttt{MatComp}: This is the algorithm from Yi et al.~\cite{yi2012crowdclustering}. 
\end{denselist}
Code for both these algorithms was provided by the authors;
we faithfully set the parameters as described in the papers.
Both these algorithms require workers to cluster random samples
of items repeatedly (recall Figure~\ref{fig:prior-work}), while ensuring that each item is assigned
the same number of times across various samples.

Note that since these algorithms do not use categorization,
we only compare the clustering phase of \orc with
these algorithms, enabling us to compare them on an equal footing.
In all executions of these algorithms, we ensured that 
the algorithms had the exact same cost as \orc,
i.e., the same number of clustering tasks assigned to workers.
We note that employing categorization would
only lead to further reduction in cost---we study the benefits
of categorization later on in this section.

\stitle{Evaluated Aspects.}
We evaluate the following aspects of \orc:
\begin{denselist}
\item How do the eventual clusterings provided by \orc
compare with the clusterings provided by other algorithms
both {\em qualitatively and quantitatively, on real-world datasets}?
\item How do the eventual clusterings provided by \orc
compare with the clusterings provided by other algorithms
both {\em qualitatively and quantitatively}, on datasets where
we can {\em control} the organizational hierarchies?
\item What is the impact, on cost and accuracy, of 
the {\em categorization} interface relative to the clustering interface?
\item What is the benefit of {\em intelligently chosen samples} 
over randomly chosen ones for \orc?
\item How does the quality of clustering vary with
the {\em number of workers}
providing clusters?
\end{denselist}

\stitle{Metrics.}
To quantitatively compare \orc with prior work, we adopt
two metrics that are also used in prior work:
{\em (a) Variation of Information (VI)}~\cite{meilua2003comparing} is an information theoretic, true distance metric used
for comparing clusterings. A VI of 0 indicates a perfect match. 
{\em (b) Normalized Mutual Information (NMI)}~\cite{cover2012elements} is a metric on $[0,1]$--- a perfect match gets a score of 1.
In addition, for our stylized dataset, we
introduce a new metric:
{\em (c) Clustering Hierarchies}, the number of hierarchies
that explain the resulting clustering returned by the algorithms.
This is calculated as the minimum number of organizational hierarchies used in assigning all items to its cluster.

\input{results}

%% file: results.tex
\newcommand{\mc}{{\tt MatComp}\xspace}
\newcommand{\cc}{{\tt CrowdClust}\xspace}

\noindent\fbox{
  \parbox{0.95\linewidth}{
  \textit{\textbf{
    \noindent How do the eventual clusterings provided by \orc
compare with the clusterings provided by other algorithms on real-world datasets?}
    }
  }
}

We compare the results of \orc
versus the other two algorithms
on the scenes and imagenet datasets.
We perform multiple runs
for each dataset (5 for scenes and 2 for imagenet); 
each run is on a subset
of $n = 115$ items
(setting $n$ as described in Section~\ref{sec:Sampling Guarantee})
--- to enable easy comparison only on clustering as opposed
to clustering plus categorization.
Each algorithm uses $5$ samples across
these $n$ items: \mc and \cc use random samples across
$n$, while \orc uses samples chosen by {\sc GenerateSample}.

\begin{figure}[hbtp]
\vspace{-20pt}
\hspace{-10pt}
\subfigure[][\orc]{
\includegraphics[scale=0.045]{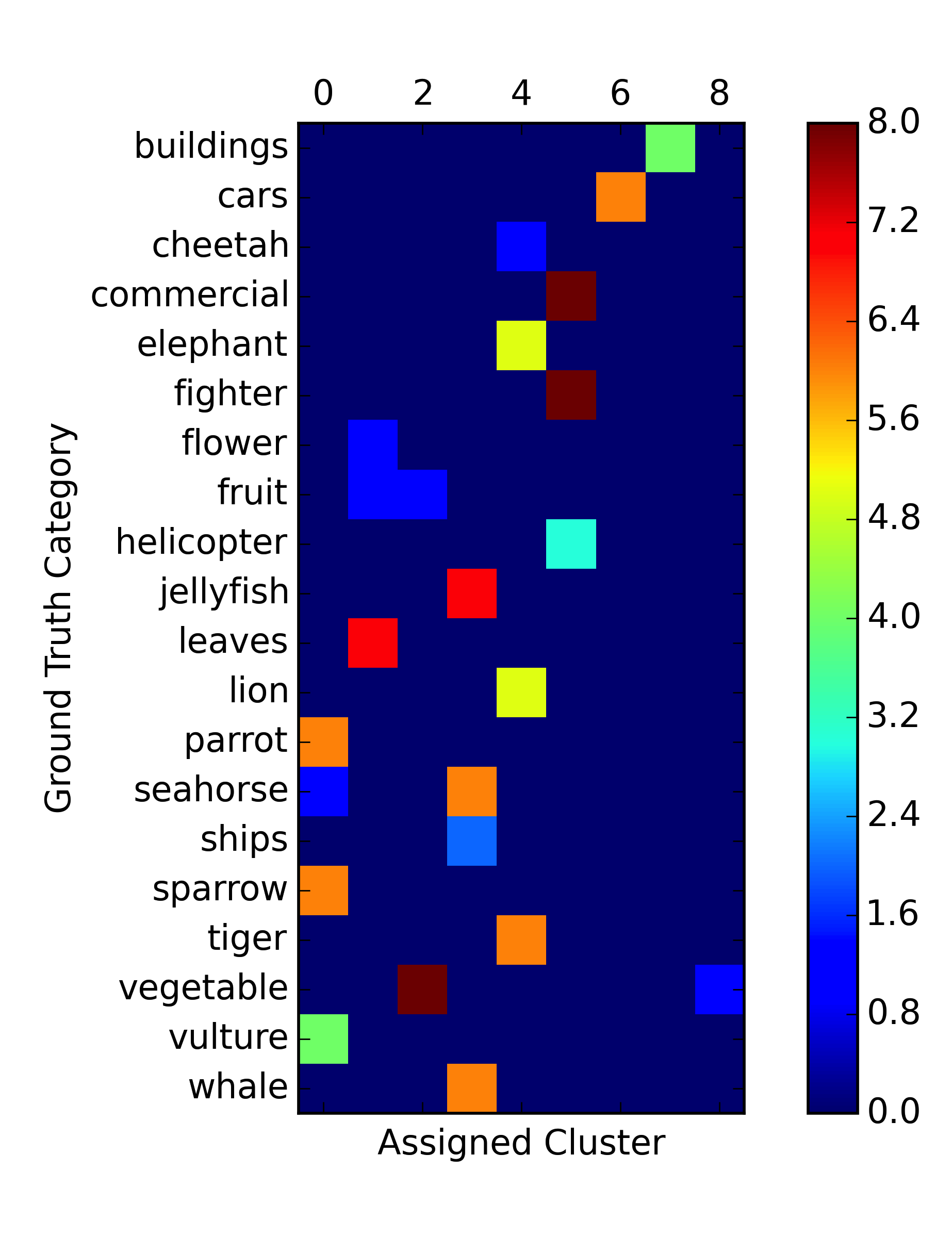}
}
\hspace{-10pt}
 \subfigure[][{\tt CrowdClust}]{
\includegraphics[scale=0.045]{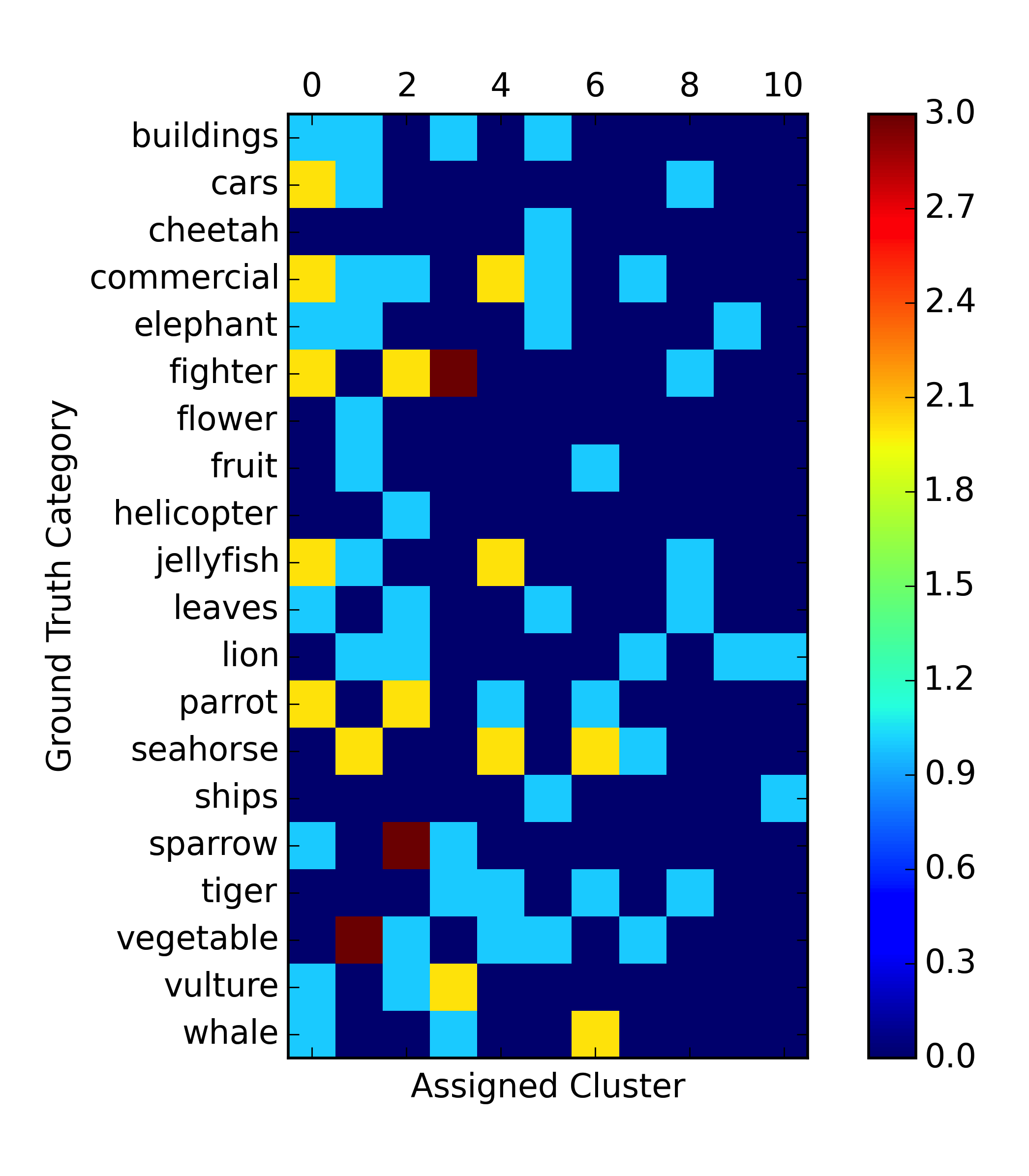}
} 
\hspace{-10pt}
\subfigure[][{\tt MatComp}]{
\includegraphics[scale=0.045]{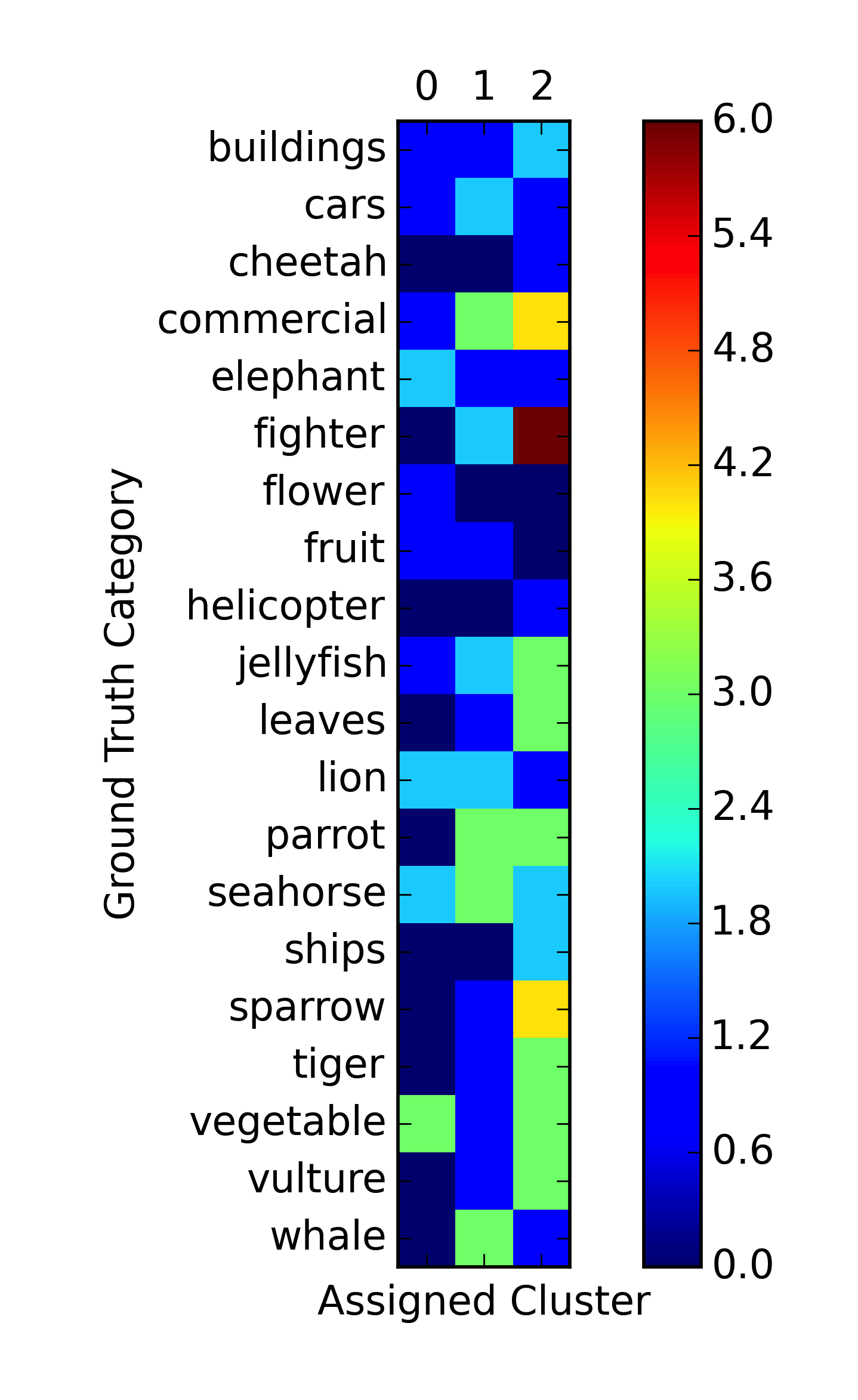}
}
%
%
\vspace{-10pt}
\caption{\label{fig:conf_matrix}Qualitative Comparison of Clusters provided by different algorithms on the ImageNet dataset}
\vspace{-10pt}
\end{figure}

We report the quantitative results in
Table~\ref{table:results},
where VI and NMI are computed for each algorithm
relative to the ground truth for the respective datasets.
We find that \orc outperforms 
both {\tt MatComp} and {\tt CrowdClust} on 
both NMI and VI (recall that smaller is better for VI and worse for NMI).
For instance, \orc has an average VI of 1.241 
across the five runs on the scenes dataset,
which is much lower than the 
1.408 of {\tt CrowdClust} and 1.635 of {\tt MatComp}
--- a 12\% and 25\% decrease respectively;
only on the fifth run does 
{\tt CrowdClust} have a better VI,
and there too the difference is $<0.02$. 
For NMI, \orc consistently performs better
than {\tt CrowdClust} and {\tt MatComp} 
in all runs.
On the imagenet dataset, 
which is a much more challenging dataset, the difference is even larger, with \orc improving over {\tt MatComp} by more than 50\% on VI, and {\tt CrowdClust} by 11\%.



\begin{table*}
\centering
\scriptsize
\begin{tabular}{ |l|l|l|l|l|l|l|l| }
\hline
& & \multicolumn{3}{ |c| }{\textbf{VI}} & \multicolumn{3}{ |c| }{\textbf{NMI}} \\
\hline
\textbf{Dataset} & \textbf{Run} \# & \orc & \texttt{CrowdClust} & \texttt{MatComp} & \orc & \texttt{CrowdClust} & \texttt{MatComp}\\
\hline
\multirow{6}{*}{Scenes} & 1 & \textbf{1.493} & 1.949 & \textbf{1.493} & \textbf{0.578} & 0.544 & \textbf{0.578}\\
 & 2 & \textbf{1.232} & 1.386 & 1.381 & \textbf{0.710} & 0.691 & 0.608\\
 & 3 & \textbf{1.171} & 1.217 & 1.423 & \textbf{0.700} & \textbf{0.700} & 0.600\\
 & 4 & \textbf{1.267} & 1.461 & 2.041 & \textbf{0.689} &  0.603 & 0.360\\ 
  & 5 & 1.040 & \textbf{1.027} & 1.837 & \textbf{0.769} & 0.761 & 0.418\\ \hline
  & \textbf{avg} & \textbf{1.2406} & 1.408 & 1.635 & \textbf{0.689} & 0.660 & 0.513\\ \hline
ImageNet & \textbf{avg} & \textbf{1.021} & 1.146 & 2.265 & \textbf{0.792} & 0.771 & 0.420\\
\hline
\end{tabular}
\vspace{-10pt}
\caption{\label{table:results}Quantitative Comparison of \orc, {\tt CrowdClust}, {\tt MatComp} on \textbf{VI} and \textbf{NMI}: For Scenes dataset, 5 runs were performed for each algorithm. The best results in each experiment are in bold.}
\vspace{-10pt}
\end{table*}


Next, we qualitatively examine the resulting clusters
for the first run of the imagenet
dataset via Figure~\ref{fig:conf_matrix}. 
We depict a confusion matrix corresponding to
each clustering algorithm,
where each ground truth category
corresponds to a row,
while each cluster in the result
corresponds to a column.
It is therefore desirable that, in each row,
there is a single column where there is non-zero presence
(i.e., the color is not blue) --- indicating
that the entire ground-truth category appears
as a whole in some cluster as opposed to being split.
On examining Figure~\ref{fig:conf_matrix},
it is clear that \cc and \mc
have much worse matrices than \orc.
For instance, the only three rows in \orc
that have non-zero presence in more than one column
are fruit, vegetable, seahorse (17/20 rows are {\em good}).
On the other hand, for \mc, all rows apart from cheetah,
flower, helicopter, and ships have non-zero presence in more than
one column --- 4/20 rows are {\em good},
and for \cc, 3/20 rows are {\em good}.
For example, the commercial category, 
referring to commercial airlines, appears
in clusters 0, 1, 2, 4, 5, 7 --- {\em seven clusters} in \cc,
and in {\em all clusters} in \mc!
Thus, there is a clear benefit to \orc in 
identifying and decomposing worker
responses across different hierarchies
and granularities,
as opposed to operating on all of them at once.
Furthermore, the clusters provided by \orc have clearly associable concepts --- cluster 4 consists of land animals; cluster 5, of aircraft,
while there are no discernible concepts corresponding to the clustering of \mc and \cc. 
Similar results hold for the other runs
on both datasets, indicating
why \orc does better than \mc and \cc on VI and NMI.
\papertext{We report on these results in our extended technical report~\cite{orchestra2015}.}



\begin{figure}[hbtp]
\vspace{-10pt}
\hspace{-10pt}
\subfigure[][\label{fig:before_categorization} before categorization]{
\includegraphics[scale=0.05]{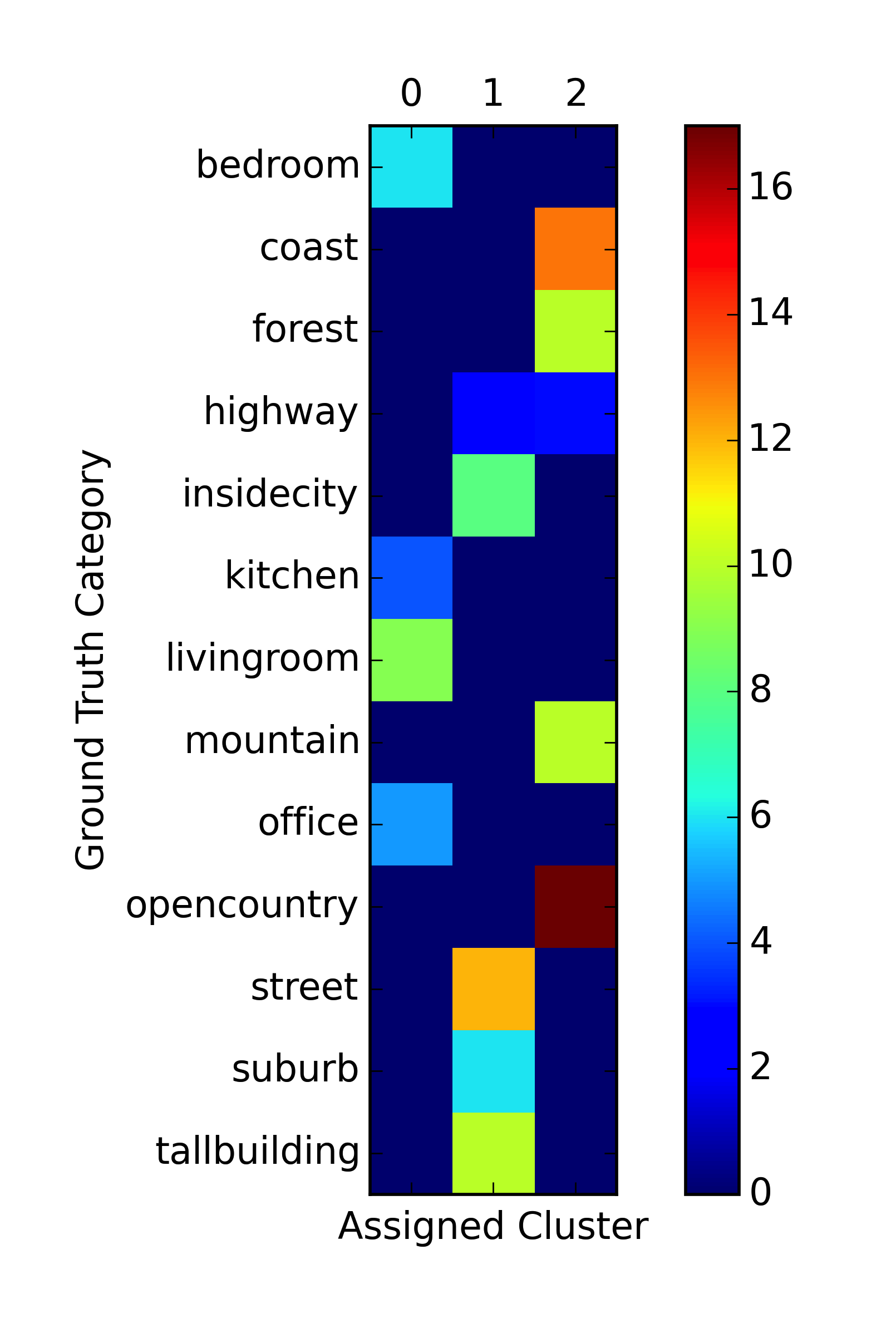}
}
\hspace{-5pt}
\subfigure[][\label{fig:orc_random} with random sampling]{
	   \includegraphics[scale=0.045]{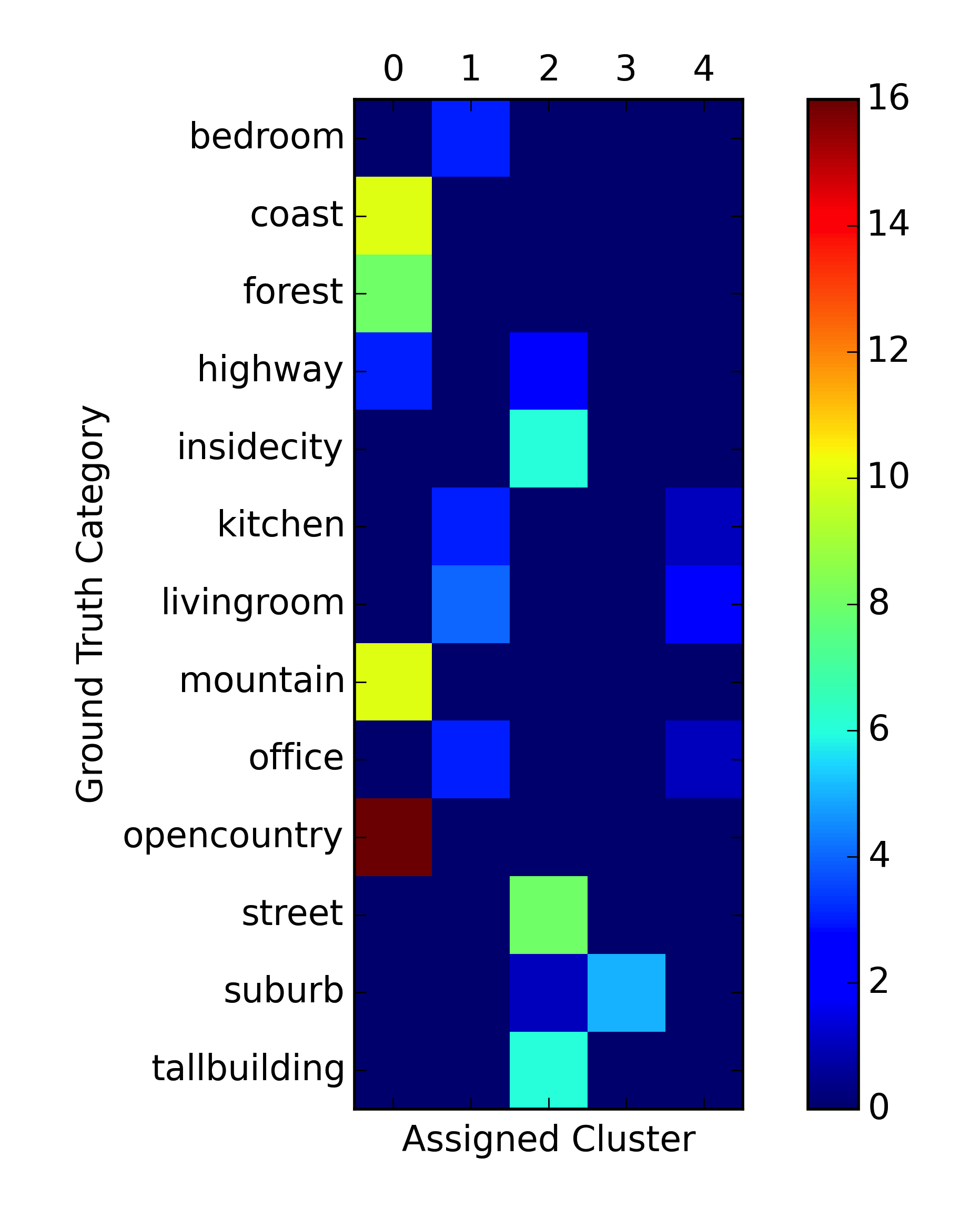}
	   }
\hspace{-5pt}
	   \subfigure[][\label{fig:after_categorization} after categorization]{
	   \includegraphics[scale=0.047]{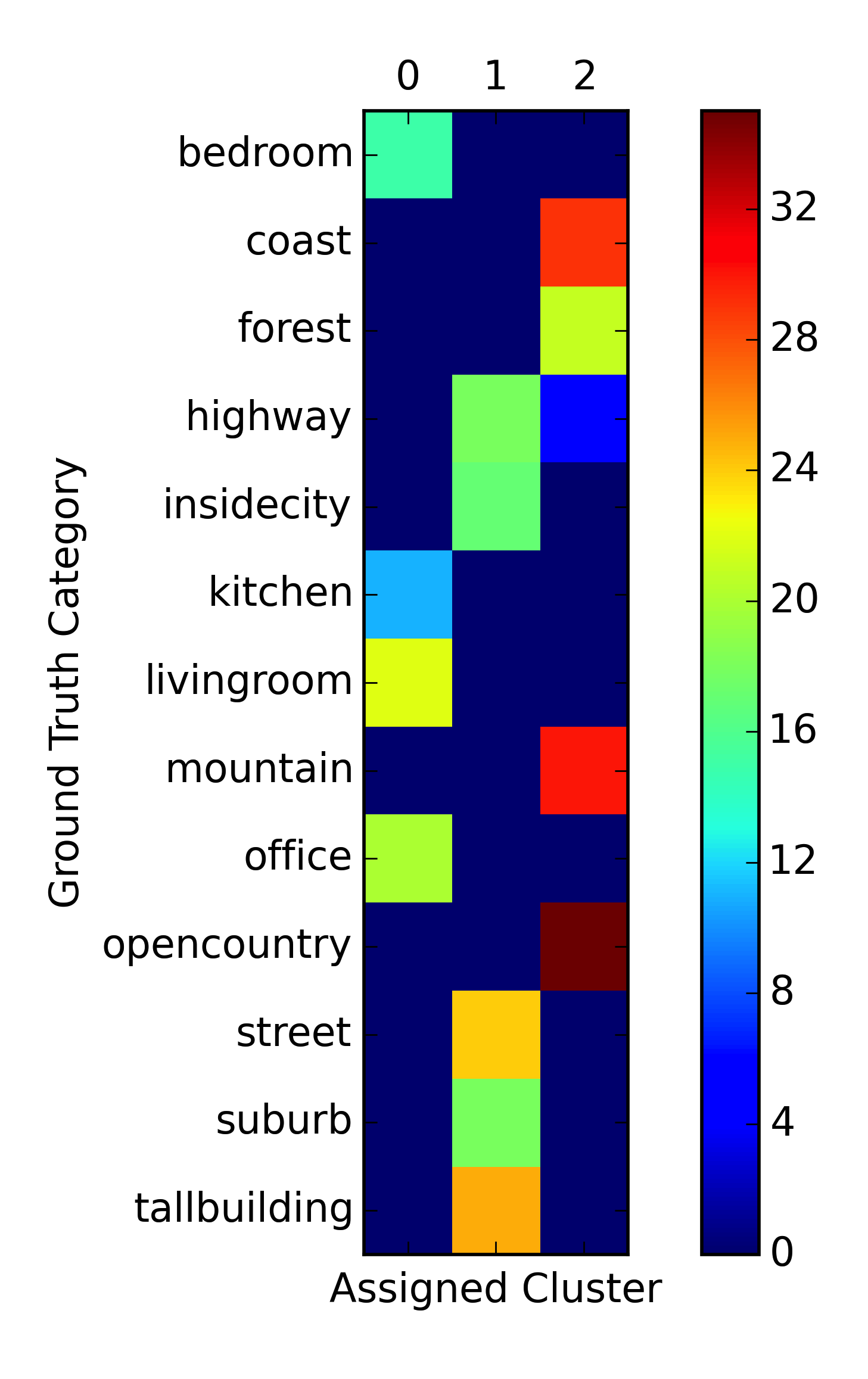}
	   }

\hspace{-10pt}
\subfigure[][\label{fig:orch_hier}\orc Hierarchy]{
        
	\tiny{\begin{tikzpicture}[%
		->,
		>=stealth,
		node distance=0.3cm,
		pil/.style={
			->,
			thick,
			shorten =2pt,}
		]
		\node (univ) {Universe};
		\node[below right=of univ] (in) {Indoor};
		\node[below left=of univ] (out){Outdoor};
		\node[below=of in] (home) {Home};		
		\node[below right=of in] (office) {Office};
		\node[below right=of out] (natural) {Natural};
		\node[below left=of out] (manmade) {City/Town};
		\draw [->] (univ) to (in);
		\draw [->] (univ) to (out);
		\draw [->] (in) to (home);
		\draw [->] (in) to (office);
		\draw [->] (out) to (natural);
		\draw [->] (out) to (manmade);
	\end{tikzpicture}}
	   }%
	   \hspace{-10pt}
\subfigure[][\label{fig:shapes_plot}Number of hierarchies for different algorithms]{\includegraphics[scale=0.10]{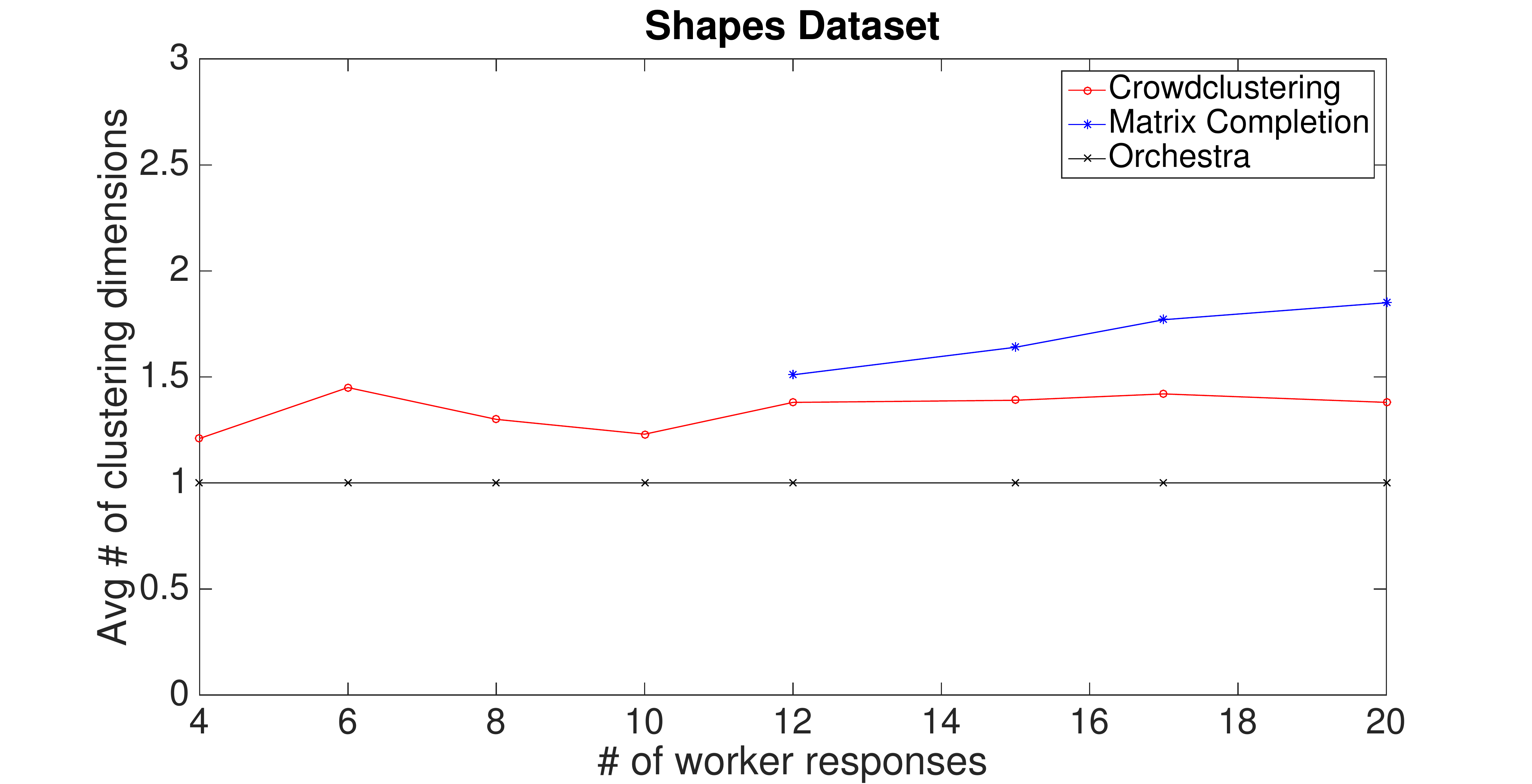}}
	   \vspace{-10pt}
	\caption{\label{fig:conf_matrix_scenes} (a -- d) Qualitative Comparison of Clusters provided by \orc on the Scenes dataset (e) Clustering Hierarchy comparison}
\end{figure}

\input{results2}

%% file: results2.tex
\noindent\fbox{
  \parbox{0.95\linewidth}{
  \textit{\textbf{
    \noindent How do the eventual clusterings provided by \orc
compare with the clusterings provided by other algorithms on a stylized dataset?}
    }
  }
}

\noindent
For this experiment, we take 20 different worker
responses for the shapes dataset:
recall that the shapes dataset is small enough
to be grouped into one single clustering task.
Thus, our \orc is restricted to the 
{\sc HierarchyConstruction} step,
and all algorithms are run on the same data,
by repeatedly taking subsets of worker responses.

For each run, we compute the number of 
hierarchies (out of \{{\sc shape}, {\sc size}, {\sc color}\})
that appear in the consensus clustering. 
Figure~\ref{fig:shapes_plot} shows the average number of organizing hierarchies vs.~the number of 
worker responses ($r$), repeated
over 100 random runs.
While \orc is always able to identify 
one dominant organizing hierarchy, 
the other algorithms tend to mix hierarchies frequently,
with the average number for \mc
being larger than \cc. 
(We were unable to run \mc for $r<12$---the spectral clustering
found no principal eigencomponents due to the matrix being low-rank.)
This is despite the fact that 
85\% of workers are actually organizing by {\sc shape} --- 
so most samplings of worker responses 
get this dominant organization.

We would also like to test all algorithms in situations when there are multiple hierarchies \emph{i.e.,} workers are equally likely to use any one of these organizations. We pick a subset of 5 responses from the 20 that we received, where 2 workers organize by {\sc shape}, 2 by {\sc color} and 1 by {\sc size}. On this data,  {\tt CrowdClust} mixes the {\sc shape} and {\sc size} organizing principles: the clustering they get is {\tt Small Shapes}, {\tt Big Triangles}, {\tt Big Rectangles}. Similarly, {\tt MatComp} is unable to come up with a consensus clustering that relies on a single hierarchy of organization. 
In contrast, \orc is able to identify a consensus clustering based on {\sc shape}. 

\input{results3}

%% file: results3.tex
\noindent\fbox{
  \parbox{0.95\linewidth}{
  \textit{\textbf{
    \noindent What is the impact, on cost and accuracy, of 
the {\em categorization} interface relative to the clustering interface?}
    }
  }
}

Earlier, we noted that the cost (per item) of the categorization phase is lower than that that in the clustering phase. 
In this section, we evaluate the quality of clusterings obtained following the categorization phase and the cost associated with it. 
For this experiment, we asked workers to categorize 175 items from the scenes dataset, organized into 5 tasks with 35 items each. 
We used 10 images from each consensus cluster found by \orc after one run on the scenes dataset, shown in Figure~\ref{fig:before_categorization} as {\em pivots} for categorization. 
We asked five independent workers to answer each task; items were assigned to the cluster with the most votes. 
 
Figures~\ref{fig:before_categorization} and \ref{fig:after_categorization} show the quality of clustering before and after categorization. Essentially, the categorization stage preserves the quality of the clustering stage, \emph{i.e.}, no new ``bad'' rows---corres\-ponding to errors---are introduced. The \texttt{highway} category was split across two clusters at the end of the clustering stage. The pivots actually help the workers in placing most of the {\tt highway} images in the \texttt{outdoor-city/town} cluster, as opposed to the \texttt{outdoor-\-natural} cluster.
 
The workers were paid 10 cents per task, as opposed to 20 cents per task for clustering. Combined with the fact that only 5 worker responses were sought (as opposed to 15 for clustering), this stage has a per-item cost that is at least 1/6-th the cost in clustering stage. Further, categorization does not require different tasks to share a common kernel of items --- each item needs to appear in only one task. This significant cost saving, while retaining the quality of the clustering stage, demonstrates that categorization is a cheap, unambiguous way of clustering items, once the clusters have been discovered with reasonable confidence.

\noindent\fbox{
  \parbox{0.95\linewidth}{
  \textit{\textbf{
    \noindent What is the benefit of intelligent sampling ({\sc GenerateSample}) compared to random sampling?}
    }
  }
}

To evaluate the impact of our {\sc GenerateSample} procedure, we run our {\sc HierarchyConstruction} and {\sc MergingHierarchies} algorithms on randomly sampled samples for the scenes dataset. One such 
result is shown in Figure~\ref{fig:orc_random}. While this is similar to the clustering with {\sc GenerateSample}, note that clusters 1 and 4 have nearly the same set of items --- {\tt indoor}. This happens because the corresponding samples have no items from {\tt indoor} that are common, and hence it is not possible to determine that these sets of items should be grouped together, resulting in a near-duplicate cluster. This is undesirable behavior, motivating the need for a {\em kernel} of items.

\noindent\fbox{
  \parbox{0.95\linewidth}{
  \textit{\textbf{
    \noindent How does the quality of clustering vary with the number of workers who perform clustering?}
    }
  }
}

\begin{figure}
\centering
	   \subfigure{\label{fig:vi_numWorkers}
	   \includegraphics[scale=0.15]{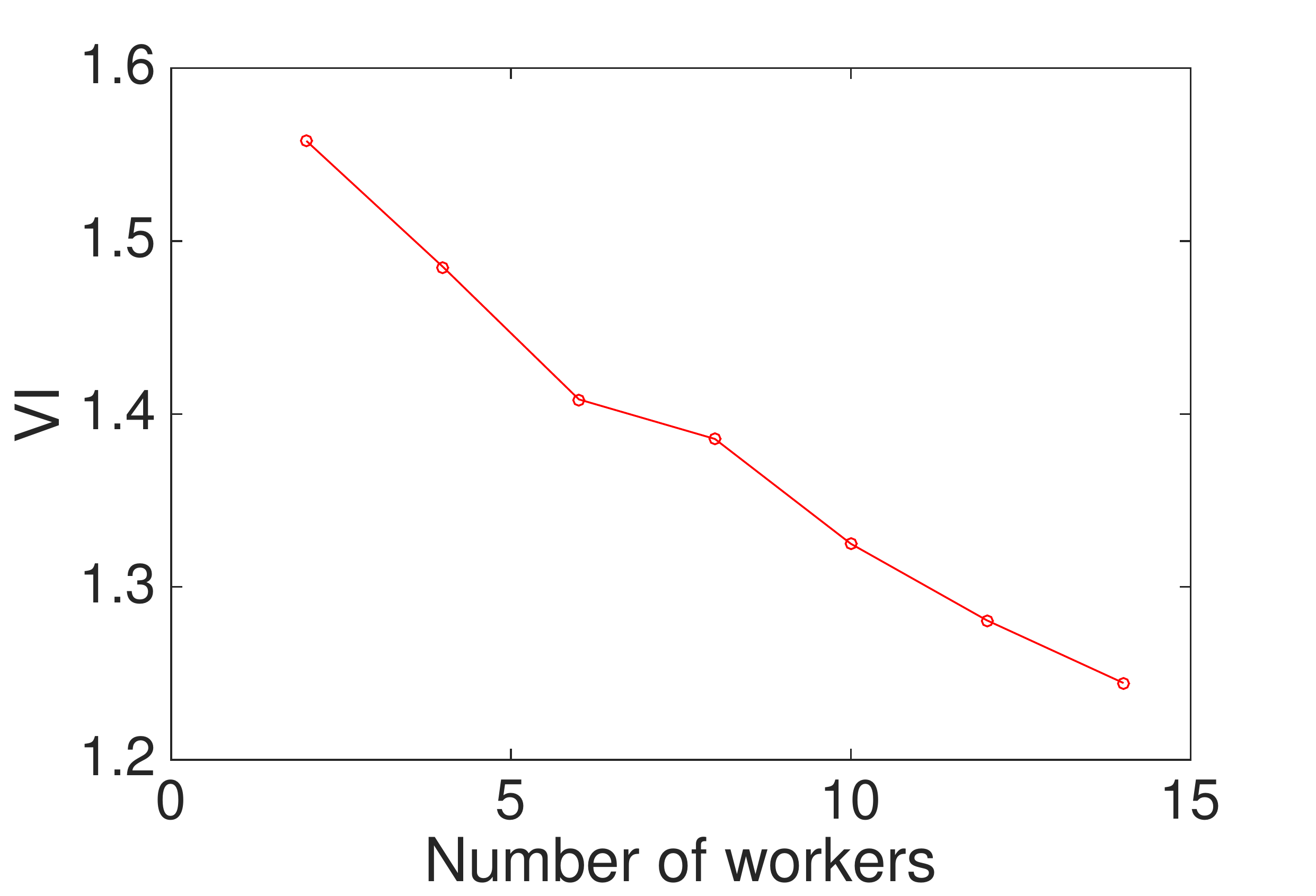}
	   }
	   \subfigure{\label{fig:nmi_numWorkers}
	   \includegraphics[scale=0.17]{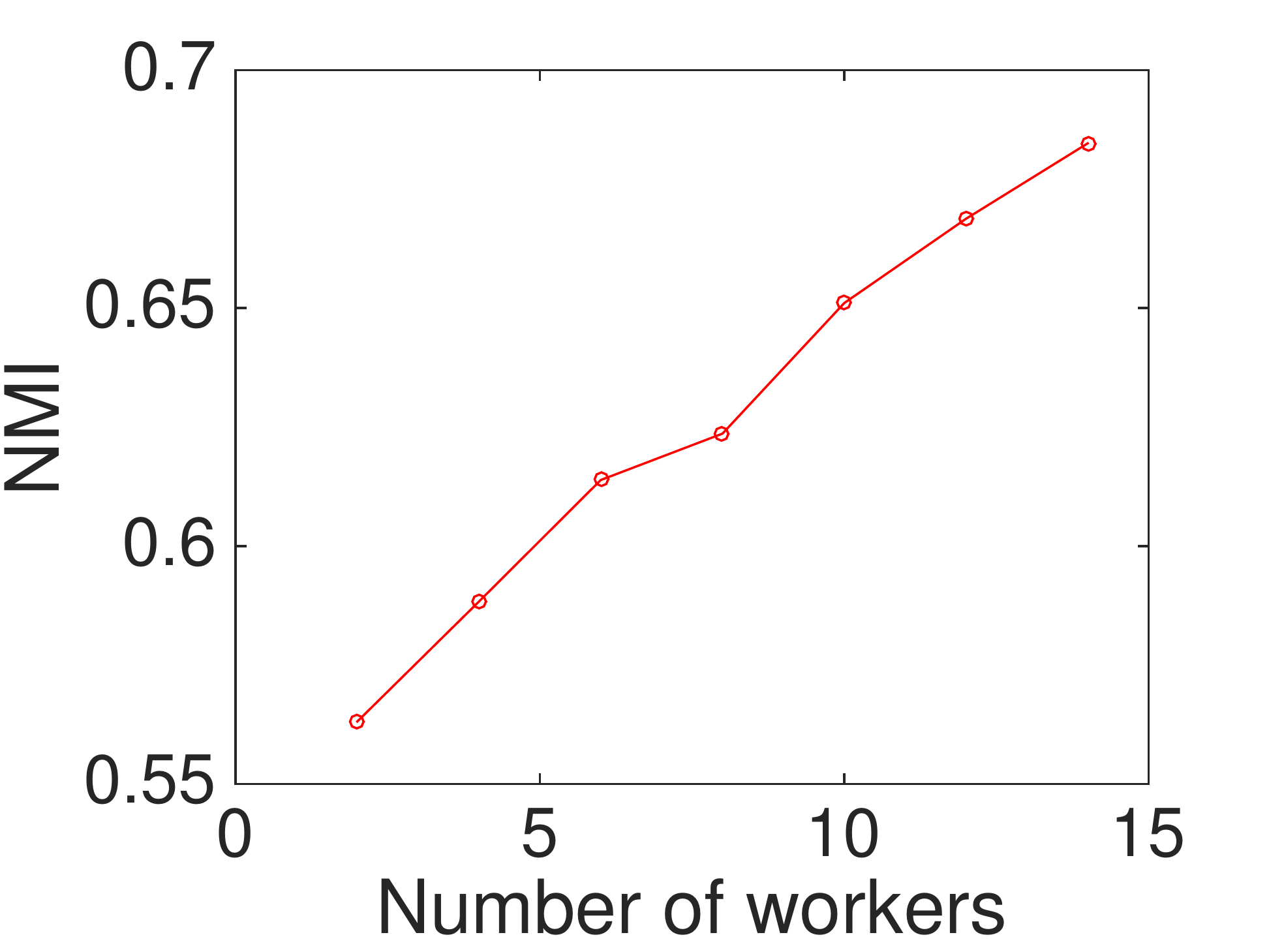}
	   }
     \vspace{-10pt}
	\caption{(a) VI and (b) NMI with varying number of worker responses}
\vspace{-15pt}
\end{figure}

To evaluate this, we plot the average VI and NMI for all runs of scenes dataset in Figures~\ref{fig:vi_numWorkers} and \ref{fig:nmi_numWorkers}. Different samples of responses are taken 50 times, and the results are averaged over these 50 trials. The performance of \orc degrades gracefully as the number of worker responses is decreased. Note that even with 6 responses, our performance (in terms of VI) is better than {\tt CrowdClust} with 15 responses (refer to Table~\ref{table:results}). Further, \orc is able to outperform {\tt MatComp} on both VI and NMI with just 2 worker responses. 
\papertext{We report on these results further in~\cite{orchestra2015}.}
 
The above results demonstrate that while more worker responses help in getting better clusterings, \orc is still able to outperform existing algorithms with fewer number of responses.

%% file: related_work.tex
\section{Related Work}\label{sec:related}

Our work is related to prior work on crowd clustering, taxonomy generation,
as well as other work on crowdsourced algorithms.

\stitle{Crowd-Based Clustering.}
Our work is most closely related to the prior work on crowd-powered clustering
via a matrix completion approach, including~\cite{gomes2011crowdclustering,yi2012semi,yi2012crowdclustering}.
In all these papers, worker clusterings are performed on randomly selected 
sets of items. 
Then, the results of worker clusterings are interpreted as pairwise comparisons:
for example, if a worker placed items a, b, c in one cluster, 
then this is interpreted as three pairwise comparisons,
between a and b, b and c, and c and a. 
Subsequently, matrix completion techniques are applied to
infer the missing entries in the matrix.
We identify multiple ways this line of work fails to take into account the complexity
of crowd-powered organization:
{\em (a)} Mixing of hierarchies and frontiers: since these papers do
not interpret different worker responses as being derived from different
hierarchies or frontiers within a hierarchy, they tend to provide
clusterings that mix hierarchies and mix frontiers within a hierarchy,
leading to poor organization.
\orc, on the other hand, carefully treats distinct hierarchies
as well as frontiers within a hierarchy. 
{\em (b)} 
Random samples of items: unlike \orc, which uses 
intelligently chosen samples of items, these papers use random samples
of items.
{\em (c)} Loss of information: since these papers interpret worker clusterings as pairwise information within
the matrix, they lose valuable information,
as opposed to \orc, which operates on clusters as a whole. 
{\em (d)}  No categorization: since the \orc approach identifies the 
consensus hierarchy, this hierarchy can be leveraged to subsequently categorize the remaining
items, providing further cost savings. 
None of these papers perform categorization to further save costs.

There has been other work on variants of clustering:
Heikinheimo and Ukkonen~\cite{heikinheimo2013crowd} describe the {\sc Crowd-Median} algorithm
whose goal is to compute centroids, as opposed to identifying clusterings.
For example, as soon as they locate {\em some} representative object, they
can stop, instead of having to organize all the objects, like in our case. 
Further, they do not explicitly capture different perspectives of workers, limiting
the applicability in practice.
Davidson et al~\cite{davidson2013using}
provide theoretical guarantees for aggregation ({\sf GROUP BY}) queries, 
where workers are asked to answer questions of the form \emph{"are a and b of the same type"}. 
This paper makes a simplifying assumption that there is a correct answer
(\emph{i.e.}, there is a ground truth collection of types),
with workers answering incorrectly with a fixed error probability.
\orc uses a more general question type (i.e., cluster a collection of objects)
since it provides more context,
and also does not make the same assumptions about worker answer correctness. 
\cite{yue2014personalized} propose a \emph{collaborative clustering} scheme where they discover user preferences for clustering as opposed to identifying a consensus clustering of the data, as we do.

\stitle{Crowd-Based Hierarchy Building.}
A variety of papers use the crowd for hierarchy construction:
Chilton et al.~and Bragg et al.~\cite{chilton2013cascade,bragg2013crowdsourcing} 
use text labels and filtering on the labels to create a hierarchy
while Sun et al.~and Karampinas et al.~\cite{sun2015building,karampinas2012crowdsourcing}
ask the crowd for pairwise ancestor descendant relationships, 
also demonstrating that identifying the optimal set of ancestor
descendant questions is {\sc NP-Hard}. 
While hierarchy construction could, in principle, be used
as a precursor to clustering or organization, 
none of these papers take into account different organizational
principles (i.e., the existence of many hierarchies); it remains to be seen if
hierarchy construction can be improved by taking into account our techniques
for identifying organizational principles. 
At the same time, it would be interesting to extend our algorithms
to generate a complete hierarchy on the set of clustered items.



\stitle{Other Crowdsourcing or Active Learning Work.}
Past work on active learning has utilized human workers to provide constraints for automated clustering algorithms. This work relies on human competence in making judgments for ambiguous image pairs, rather than using humans expertise in organizing data into clusters. Biswas et al.~\cite{biswas2014active} obtain hard pairwise constraints in a crowdsourced setting by asking targeted questions related to an item pair. Lad et al.~\cite{lad2014interactively} ask humans to provide attribute-based explanations, rather than pairwise constraints, and opt to use these as soft constraints. Neither work allows humans to explicitly cluster data.

Other work focuses on learning a embedding of the data using crowd workers, and then clustering in this latent space using a standard clustering algorithm. The disadvantage of this approach is that it mashes together worker responses, and loses rich information that can be extracted from workers. Wilber et al.~\cite{wilber2015learning} create concept embeddings by combining human experts with automation. Tamuz et al.~\cite{tamuz2011adaptively} learn a `crowd kernel', which embeds items into a Euclidean space. Neither work explores how to cluster this embedding effectively, to extract different organizational hierarchies. 

Prior work on categorization does not attempt to discover organizational principles, instead presenting a predefined organization to workers, and asking them to assign items into categories. 
Both papers in this space~\cite{parameswaran2011human,fan2015icrowd} use graph-based approaches to carry out categorization into a taxonomy of concepts. Our work can be considered a precursor to the algorithms described in these papers, which can be integrated into our categorization step.

Work on entity resolution (ER) can be regarded as clustering with a different objective: find clusters of homogenous (identical) items, in contrast to our setting, where the organizing principle is not clear. \cite{lee2013hybrid,whangcompare,whang2013question,vesdapunt2014crowdsourcing,wang2012crowder} are all examples of work that rely on human judgments to carry out ER, all using pairwise comparisons. 